\documentclass[floatfix,twocolumn,superscriptaddress]{revtex4}

\usepackage{xcolor}
\usepackage{mathrsfs}
\usepackage{graphicx}
\usepackage[english]{babel}
\usepackage{amsmath}
\usepackage{amssymb}
\usepackage{booktabs} 
\usepackage{array} 
\usepackage{tabularx}
\usepackage{makecell} 
\usepackage{relsize}
\usepackage{CJK}
\usepackage{dsfont}
\usepackage{bm}
\usepackage{xcolor}

\newcommand{\me}{\mathrm{e}}
\newcommand{\mi}{\mathrm{i}}

\allowdisplaybreaks
\begin{document}

\title{Machine learning of the Ising model on a spherical Fibonacci lattice}

\author{Zheng Zhou}
\affiliation{School of Physics, Southeast University, Jiulonghu Campus, Nanjing 211189, China}

\author{Chen-Hui Song}
\affiliation{Tsinghua Shenzhen International Graduate School, Tsinghua University, Shenzhen 518055, China}

\author{Xu-Yang Hou}
\email{houxuyangwow@seu.edu.cn}
\affiliation{School of Physics, Southeast University, Jiulonghu Campus, Nanjing 211189, China}

\author{Hao Guo}
\email{guohao.ph@seu.edu.cn}
\affiliation{School of Physics, Southeast University, Jiulonghu Campus, Nanjing 211189, China}
\affiliation{Hefei National Laboratory, Hefei 230088, China}

  \begin{abstract}
We investigate the Ising model on a spherical surface, utilizing a Fibonacci lattice to approximate uniform coverage. This setup poses challenges in achieving consistent lattice distribution across the sphere for comparison with planar models. We employ Monte Carlo simulations, principal component analysis (PCA), graph convolutional networks (GCNs) to study spin configurations across a range of temperatures and to determine phase transition temperatures. The Fibonacci lattice, despite its uniformity, contains irregular sites that influence spin behavior. In the ferromagnetic case, sites with fewer neighbors exhibit a higher tendency for spin flips at low temperatures, though this effect weakens as temperature increases, leading to a higher phase transition temperature than in the planar Ising model. In the antiferromagnetic case, lattice irregularities induce geometric frustration, resulting in highly degenerate ground states and the phase transition temperature lower than the planar square lattice.
Phase transition temperatures are derived through specific heat, magnetic susceptibility analysis and GCNs predictions, yielding $T_c$ values for both ferromagnetic and antiferromagnetic scenarios. This work emphasizes the impact of the Fibonacci lattice's geometric properties-namely curvature and connectivity-on spin interactions in non-planar systems, with relevance to microgravity environments.
  \end{abstract}
\maketitle

 \section{Introduction}

The Ising model is perhaps the simplest statistical spin model, yet it exhibits a wealth of physical phenomena and plays a crucial role across multiple fields of physics. Experimentally, the Ising model can be realized using cold-atom quantum simulators \cite{PhysRevLett.25.443,PhysRevA.77.051601,NatureQI16,WOS:000416520400033,WOS:000327944600013}. Recently, the rapid advancement of space-based technology has driven experimental efforts to confine ultracold atoms on surfaces of various shapes in microgravity \cite{SphericalBECnpj19,SphericalSFPRL20,carollo2022observation,PhysRevResearch.4.013122,PhysRevLett.125.010402,PhysRevLett.123.160403,WOS:000502992700002, PhysRevA.103.053306, doi:10.1126/science.1189164, WOS:000447807100054}. Among these adjustable shapes, the sphere or spherical bubble trap has garnered significant research interest, with numerous studies dedicated to exploring its geometric effects on cold atomic gases. In this paper, we focus on the Ising model confined to a spherical surface. Given the fundamental nature of the Ising model, its analysis is crucial for understanding the physical characteristics of spherical atomic gases.

The first challenge of this problem is to cover the spherical surface as uniformly as possible. Only in this way can we reliably compare the results with those of the planar square lattice. This requirement excludes the traditional latitude-longitude lattice, as the density of sites near the poles is significantly higher than anywhere else. The most suitable candidate is the Fibonacci lattice, which is essentially the most uniform spherical lattice.
It is important to note that the Fibonacci lattice is not perfectly uniform, which complicates the analytical study of the Ising model on the Fibonacci sphere. However, by utilizing Monte Carlo (MC) simulations and machine learning techniques, we can determine the spin configurations at various temperatures and identify the phase-transition temperature of the spherical Ising model. 

The second challenge arises from the irregular nature of the Fibonacci lattice, which lacks an image-like structure. Machine learning techniques have been widely used to identify phase transitions in statistical physics models, such as the Ising model. Notably, Carrasquilla and Melko \cite{MLNP17} demonstrated the effectiveness of supervised learning with convolutional neural networks (CNNs) in classifying ordered and disordered phases in the two-dimensional square-lattice Ising model, successfully determining the critical temperature from labeled configurations. However, this approach cannot be directly applied to the Fibonacci lattice due to its non-uniform geometry. To address this, we extend the paradigm of machine learning for phase transition detection to the irregular geometry of the spherical Fibonacci lattice. Specifically, we employ graph convolutional networks (GCNs), a natural generalization of CNNs for non-Euclidean structures.

In our previous work, we applied the GCNs approach to study the XY model on the spherical Fibonacci lattice \cite{PhysRevResearch.4.023005}, yielding intriguing results about how vortex distribution is significantly influenced by spherical topology. This demonstrates that the method is well suited for handling continuous spin systems and complex non-Euclidean geometries.

Recently, a study focused on the ferromagnetic Ising model on a Fibonacci-triangulated sphere found that the model exhibits a critical temperature slightly lower than that of a planar triangular lattice \cite{SIsing23}. In this paper, we will concentrate on the mostly quadrangulated Fibonacci lattice for both ferromagnetic and antiferromagnetic Ising models, where neighbor interactions are determined by a cutoff radius $r_c$. Through the application of specific heat analysis and GCNs, we determined the phase transition temperatures for both situations, highlighting the influence of geometric properties on spin interactions.

The rest of this paper is organized as follows: In Section \ref{II}, we construct the spherical Fibonacci lattice points, ensuring that the number of lattice sites with four nearest neighbors is maximized by selecting an appropriate spin-interaction cutoff radius $r_c$. Sections \ref{III} and \ref{IV} present simulations of the spherical ferromagnetic and antiferromagnetic Ising models, respectively, using MC algorithms, PCA and GCNs. These sections include graphs illustrating spin configurations, energy, specific heat, magnetization and magnetic susceptibility at various temperatures, as well as the phase transition temperatures for both models. Finally, Section \ref{V} concludes the paper.

\section{Spherical Ising Model}\label{II}

The Ising model on a 2D lattice is described by the Hamiltonian:
\begin{equation}\label{e1}
H_{\left\{s_{i}\right\}}=-J \sum_{\langle i, j\rangle} s_{i} s_{j}-h \sum_{i} s_{i},
\end{equation}
where $s_i$ represents the spin at site $i$, $J$ is the interaction strength and $h$ is the external magnetic field. The first summation runs over all pairs of adjacent spins, with each pair being counted only once. In this paper, we will focus on the case with $h=0$ for convenience.
Ising models can be classified according to the sign of $J$: If $J>0$, the interaction is called ferromagnetic; if $J<0$, it is antiferromagnetic.
For irregular lattices, the distance between adjacent spins may not be a constant. Therefore, it is crucial to carefully define the ``nearest-neighbor sites'', which will be discussed in detail later.

\begin{figure}[ht]
\centering
\includegraphics[width=1.24in]{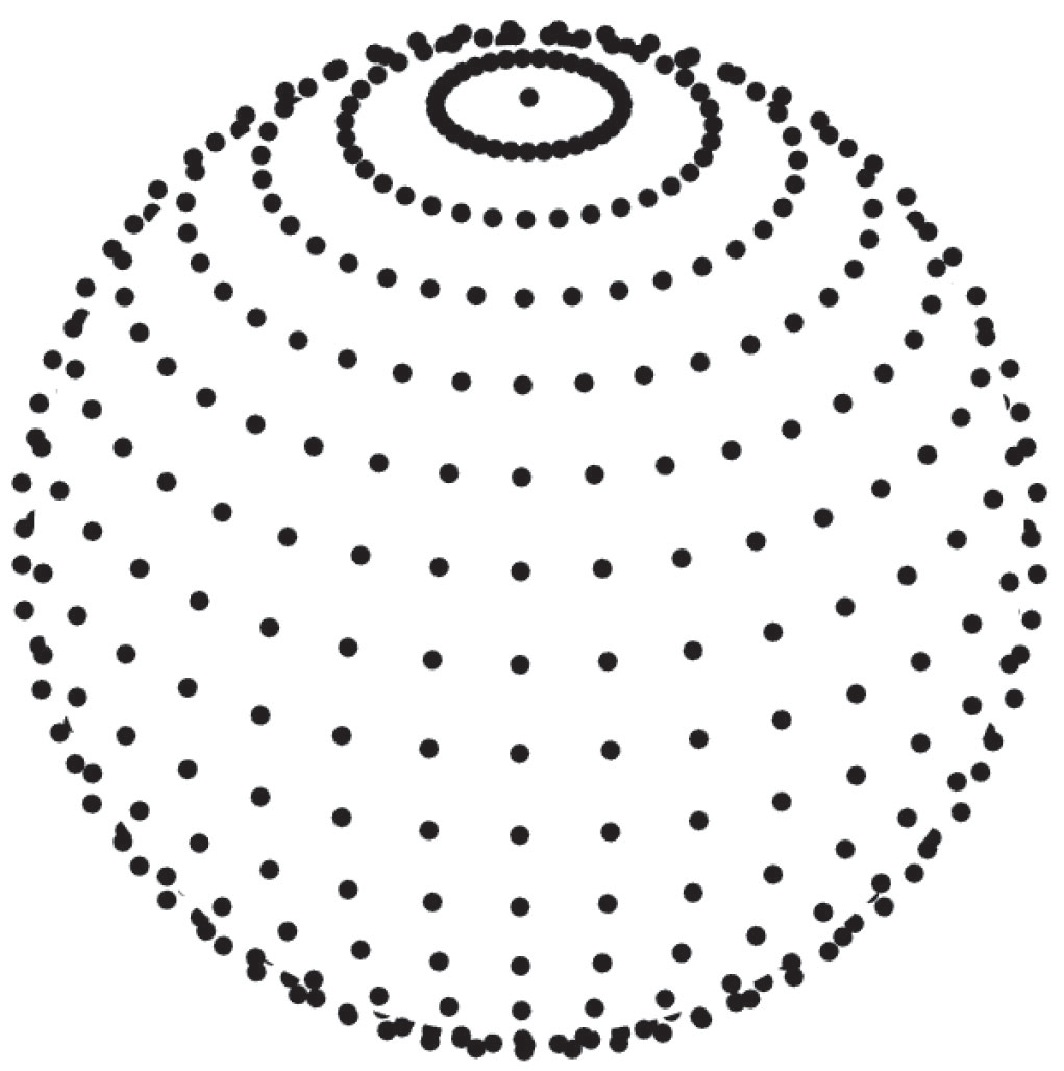}
\hspace{0.2in}
\includegraphics[width=1.24in]{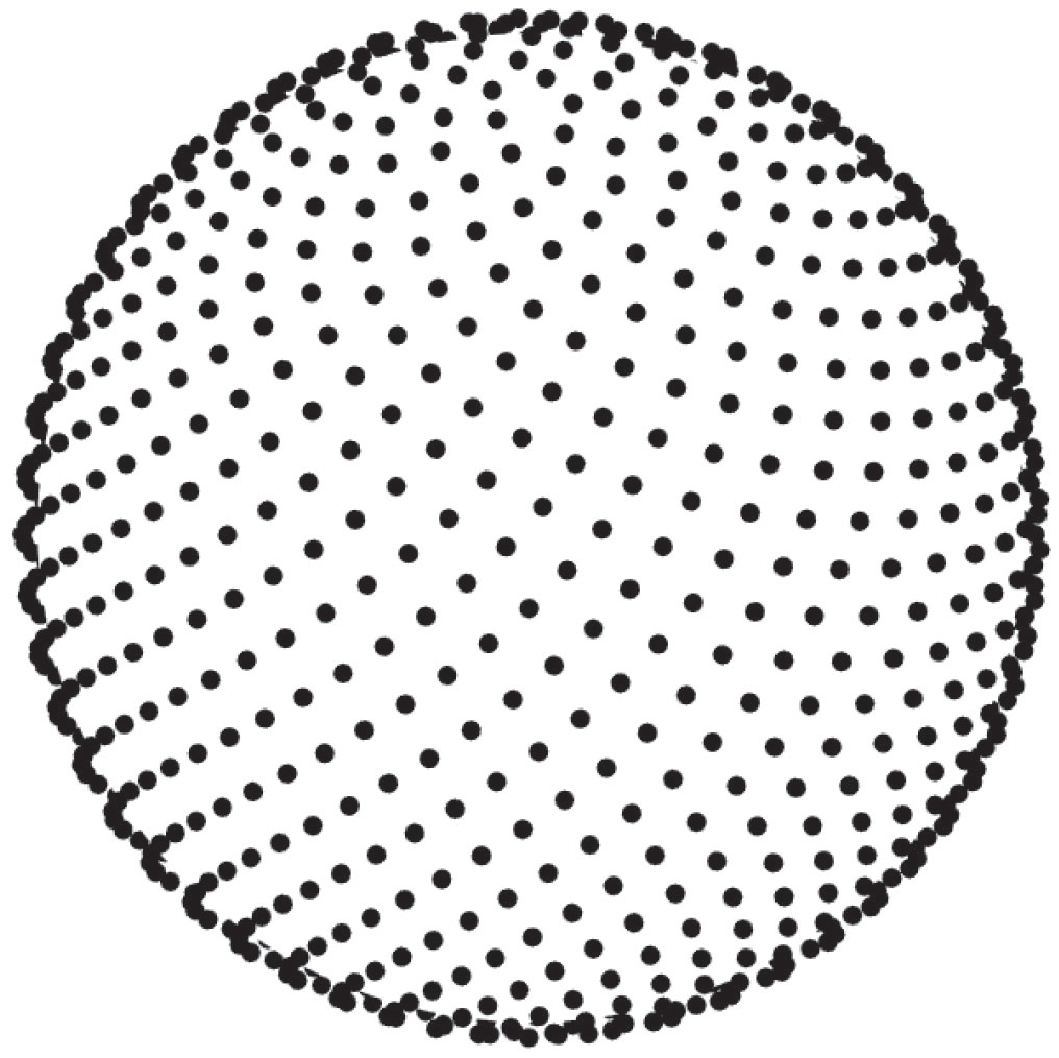}
\caption{A comparison of the uniformity between two types of lattices of $N=1000$: the latitude-longitude lattice on the left and the Fibonacci lattice on the right.}	\label{lattice}
\end{figure}

\begin{figure}[ht]
\centering
\includegraphics[width=1.24in]{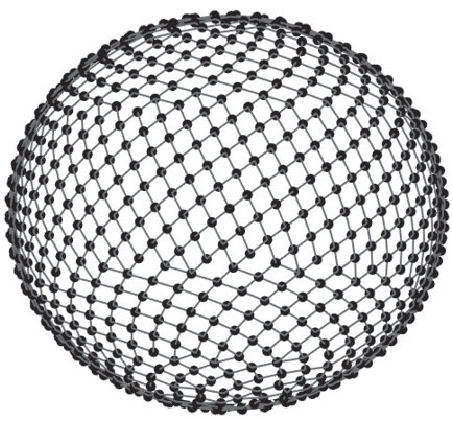}
\hspace{0.2in}
\includegraphics[width=1.28in]{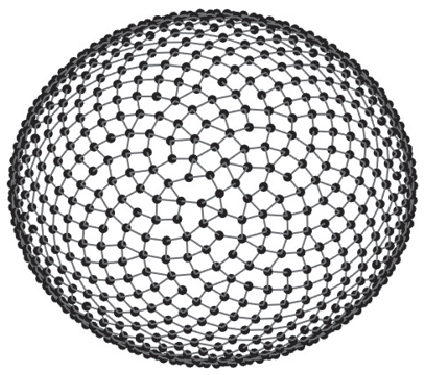}
\caption{Perspectives of a $N=1000$ Fibonacci lattice from different directions.}	\label{connect}
\end{figure}

The spins of a Fibonacci Ising model are located on a Fibonacci lattice on a spherical surface. The position of each lattice site in Cartesian coordinates is given by:  \cite{WOS:000273034500003, Hannay_2004, WOS:000363671200030}
\begin{align}\label{e2}
x_{i} & =\sqrt{R^{2}-z_{i}^{2}} \cos (2  \pi \mi\phi), \quad y_{i}=\sqrt{R^{2}-z_{i}^{2}} \sin (2  \pi \mi\phi), \notag \\
z_{i} & =R\left(\frac{2 i-1}{N}-1\right),
\end{align}
where the notation $i=1,2, \cdots, N$ denotes the index of a lattice point, $R$ represents the radius of the sphere, and $\phi=\frac{\sqrt{5}-1}{2}$ is the golden ratio.
Fig.\ref{lattice} compares the traditional latitude-longitude lattice with the Fibonacci lattice. Evidently, the latter is considerably more uniform than the former. Next, to compare with the properties of the Ising model on the planar square lattice, we quadrangulate the spherical Fibonacci lattice as extensively as possible. In essence, this means that the majority of spins interact with four nearest neighbors. To achieve this, we set a critical radius $r_c$. If the distance between two neighboring spins is less than $r_c$, an interaction is considered to exist between them.
Firstly, we consider a system with $N=1000$ lattice sites as an example.
To maximize the number of lattice sites with four nearest neighbors, we perform an exploratory analysis to determine the optimal nearest-neighbor radius, setting it to $r_c=0.1298R$. Under these conditions, 850 spins have four neighbors, 76 spins have three neighbors, and 74 spins have five neighbors. In total, the 1000 spins collectively have 3998 neighboring connections, closely approximating the structure of a square lattice with $N=1000$. We connect all ``nearest-neighbor sites'' and present a two perspectives of the quadrangulation of a $N=1000$ Fibonacci lattice from two different directions in Fig.\ref{connect}. Most sites appear ``regular'' except a few having 3 or 5 neighbors. In practical experiments, implementing a Fibonacci lattice is quite straightforward. We simply need to distribute the lattice points as evenly as possible on a spherical surface. This arrangement will naturally approximate a Fibonacci lattice in an appropriate coordinate system, as the Fibonacci lattice is fundamentally the most uniform lattice on a sphere.

\section{Ferromagnetic Ising model on a spherical Fibonacci lattice}\label{III}

\subsection{Spin configurations}

We first consider the case with $J>0$, in which neighboring sites tend to align with the same spin orientation. This is referred to as the spherical ferromagnetic Ising model.
When mapping the Ising model on a spherical Fibonacci lattice, $r_c$ serves as a cutoff range of the interaction.
We set $J=1$, $N=1000$, and $r_c/R=0.1298$. and use the MC techniques to generate the spin configurations. Specifically, a spin is randomly selected and proposed to flip, and the Metropolis criterion is applied to decide whether the flip is accepted \cite{Metropolis1953, RM4035RC, lang1973metropolis}. If the flip lowers the system energy, it is always accepted; otherwise, it is accepted with a certain acceptance probability defined as
\begin{align}
P_{\text{accept}} = \min\left(1, \me^{-\Delta E / k_B T}\right),
\end{align}
where $\Delta E$ is the energy change caused by the flip, $k_B$ is the Boltzmann constant, and $T$ is the temperature. As the temperature decreases, the probability of accepting energy-increasing flips becomes smaller, reflecting the tendency of the system to remain in low-energy states and suppress disordered configurations. In the zero temperature limit, only energy-lowering flips are allowed, and the system eventually converges to the ground state. At finite temperature, after a sufficient number of spin-flip attempts, the statistical distribution of spin configurations converges to the Boltzmann distribution,
\begin{align}
P(\{s_i\}) \propto e^{-E(\{s_i\})/k_BT}.
\end{align}
At this stage, the acceptance probability no longer changes with the number of Monte Carlo steps but instead fluctuates around a stable mean value, indicating that the system has reached a statistical steady state, i.e., thermal equilibrium.

\begin{figure}[htbp]
\centering
\includegraphics[width=1.24in]{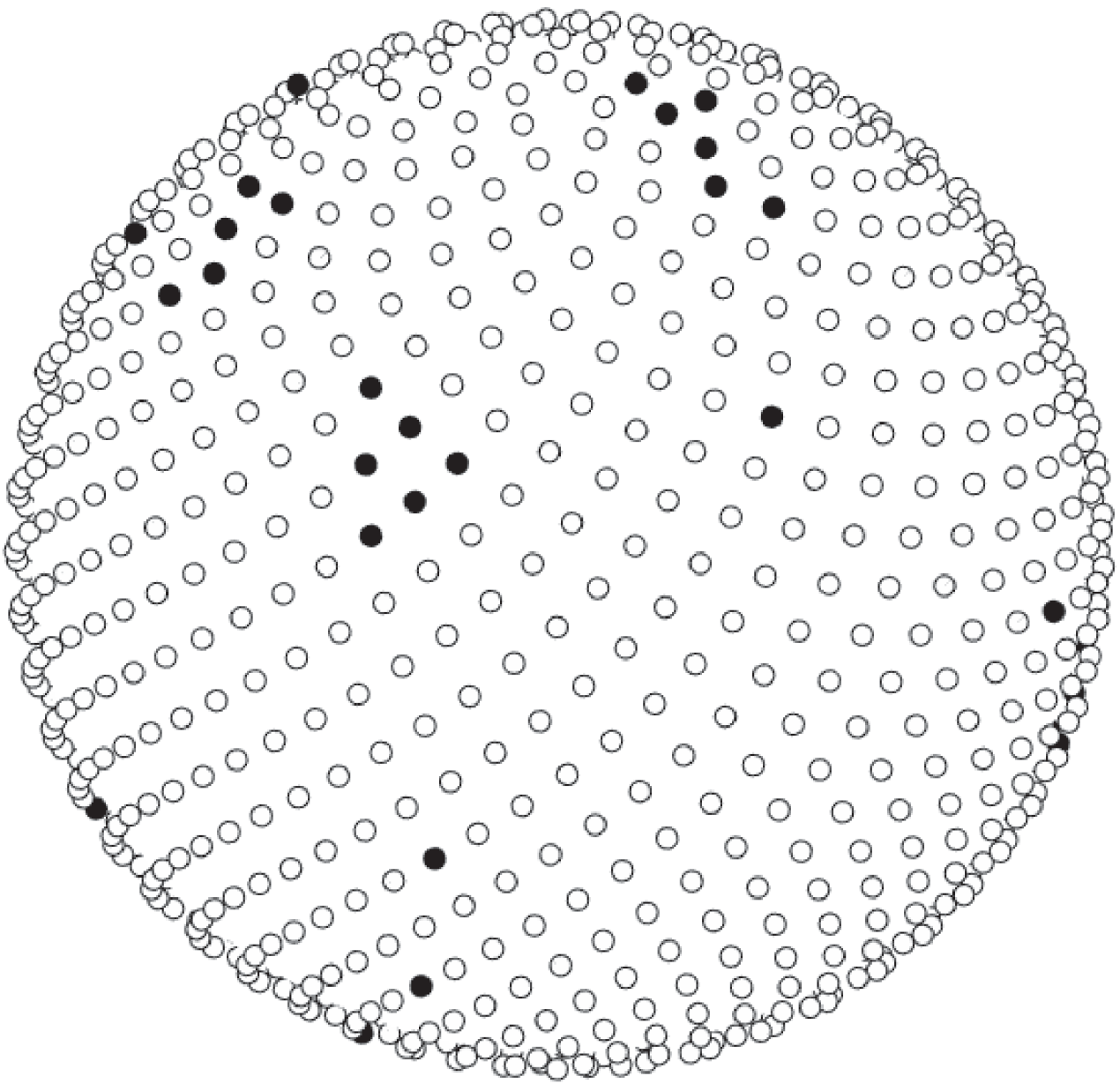}
\hspace{0.2in}	
\includegraphics[width=1.24in]{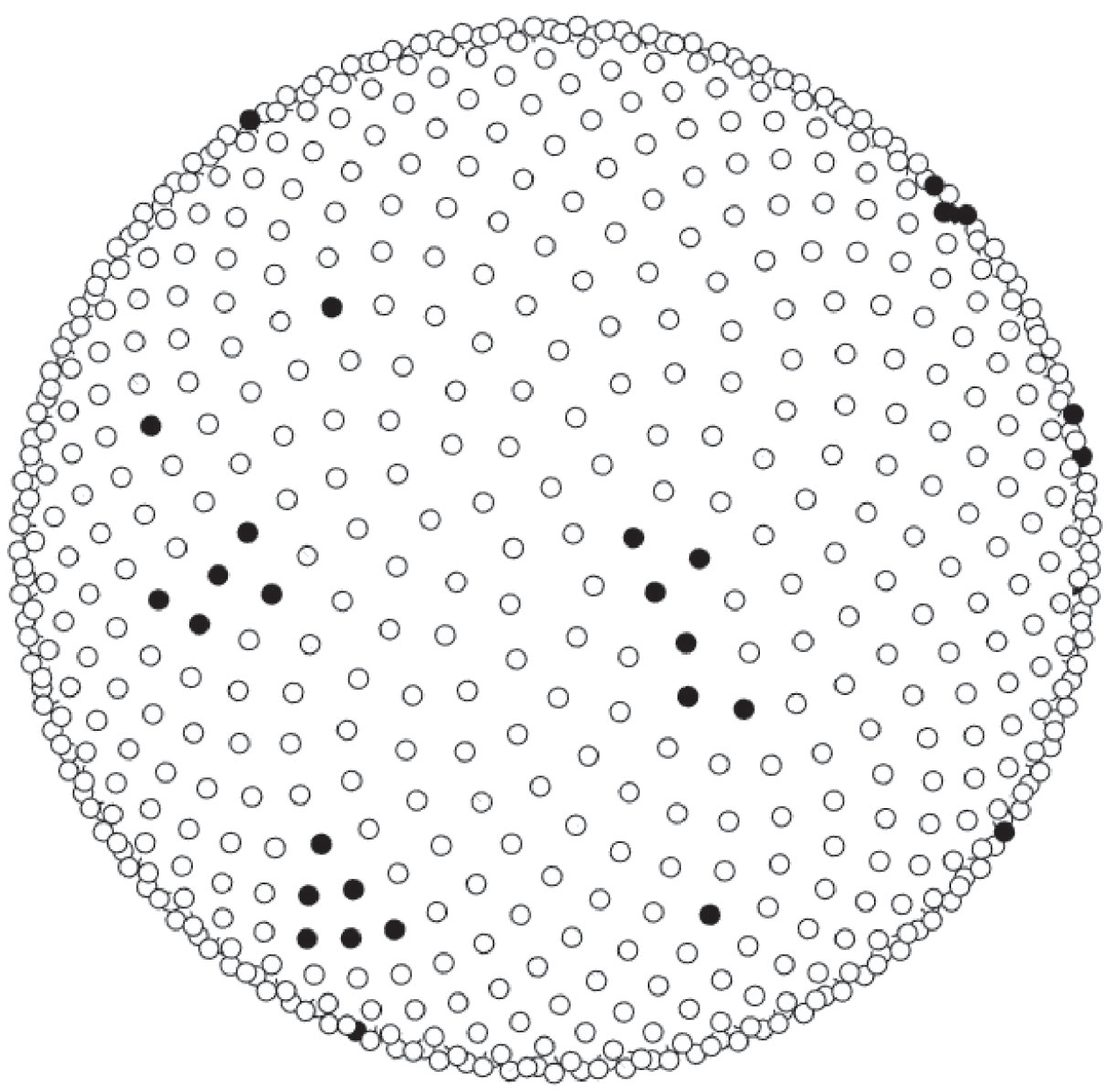}\\
\includegraphics[width=1.24in]{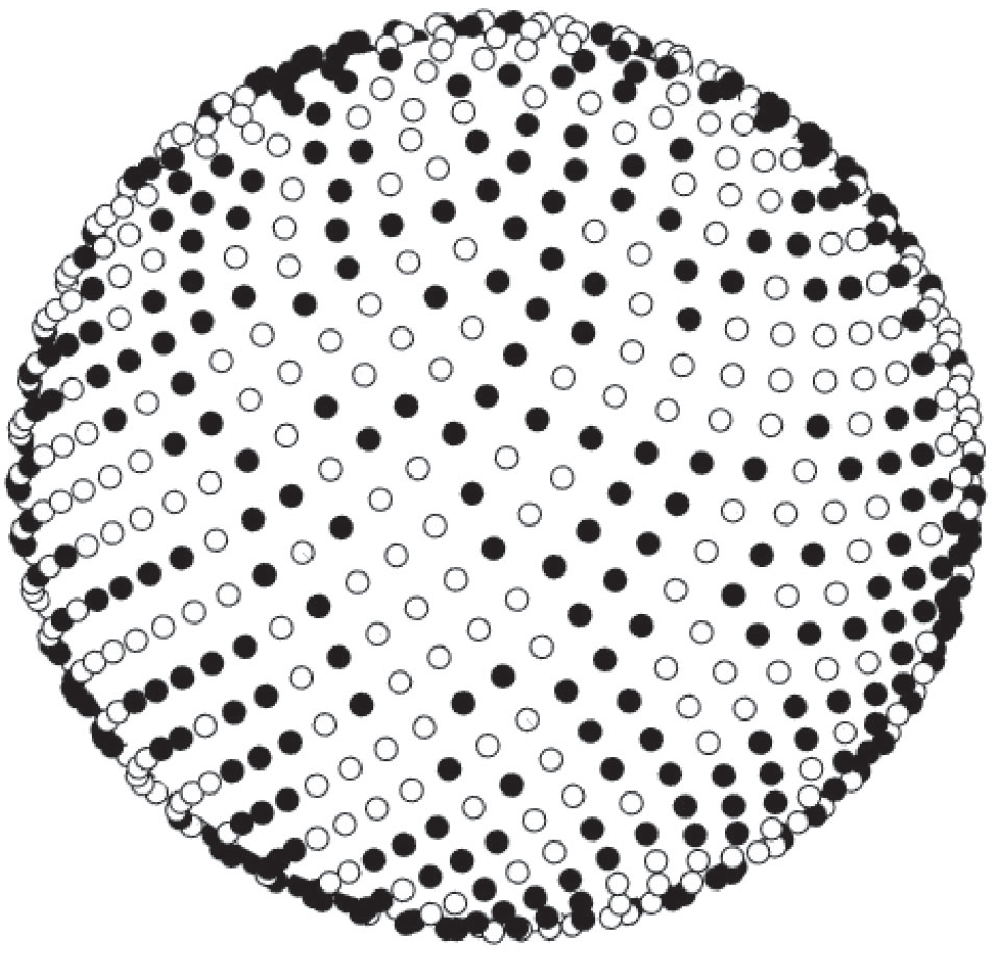}
\hspace{0.2in}	
\includegraphics[width=1.24in]{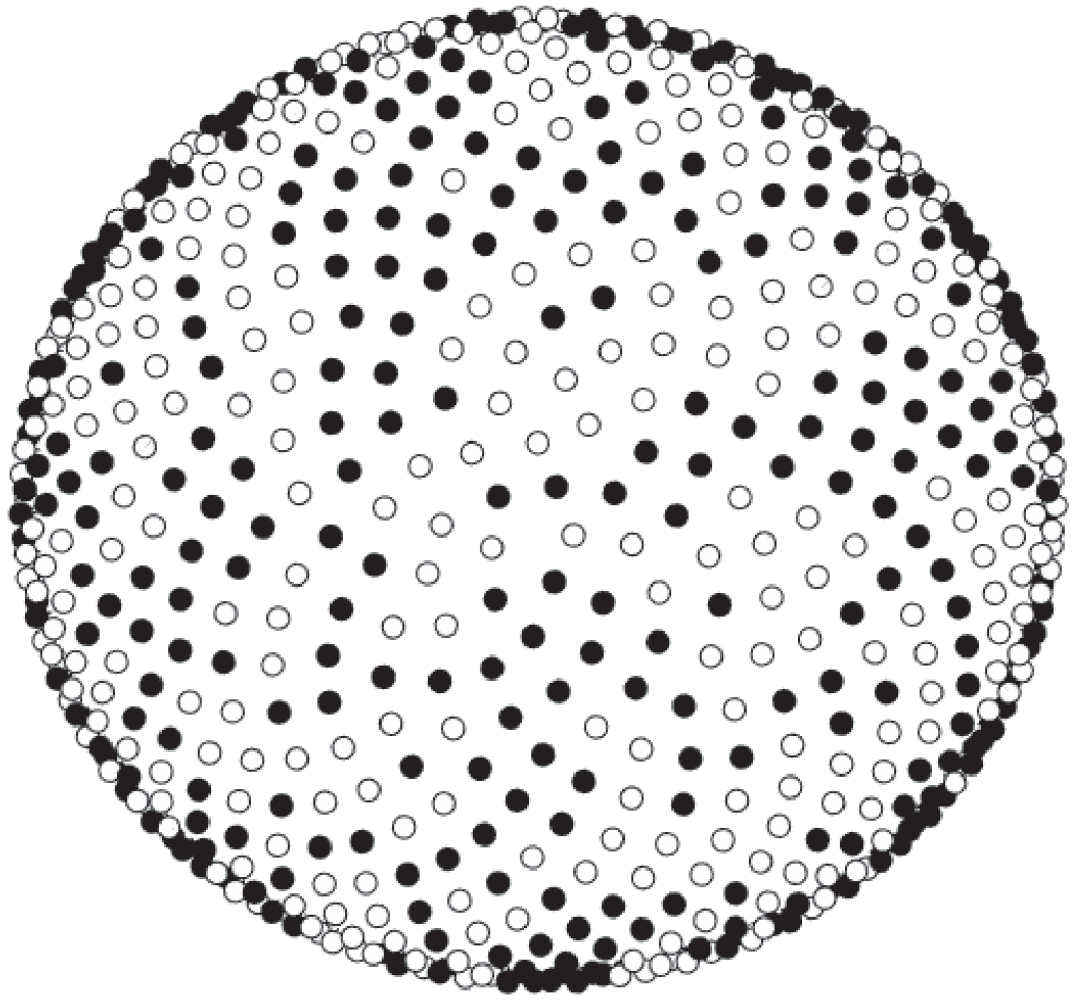}
\caption{Spin configuration of the spherical ferromagnetic Ising Model at $T/J=2.0$ (Top panel) and $T/J=8.0$ (Bottom panel):  the left panel shows the front view, while the right panel presents the top view. Up and down spins are white and black pixels. }	
\label{lowtem}
\end{figure}

At low temperatures, it is evident that all sites have the same spin orientation. As the temperature starts to increase, thermal fluctuations may cause spins at certain sites to reverse direction.
In the top panel of Fig.\ref{lowtem}, we plot the stable spin configuration at a relatively low temperature of $T/J=2.0$, where up and down spins are represented by white and black pixels, respectively. It is observed that some spins are flipped (black points), while the majority of the regions remain predominantly occupied by up spins (white points). Clearly, the system is in the ordered phase. At a very high temperature of $T/J=8.0$, a significant number of spins are flipped due to thermal fluctuations. As a result, the numbers of up and down spin sites become roughly comparable. The system is now in the disordered phase, and the corresponding spin configuration is shown in the bottom panel of Fig. \ref{lowtem}.

\begin{figure}[ht]
\centering
\includegraphics[width=1.23in]{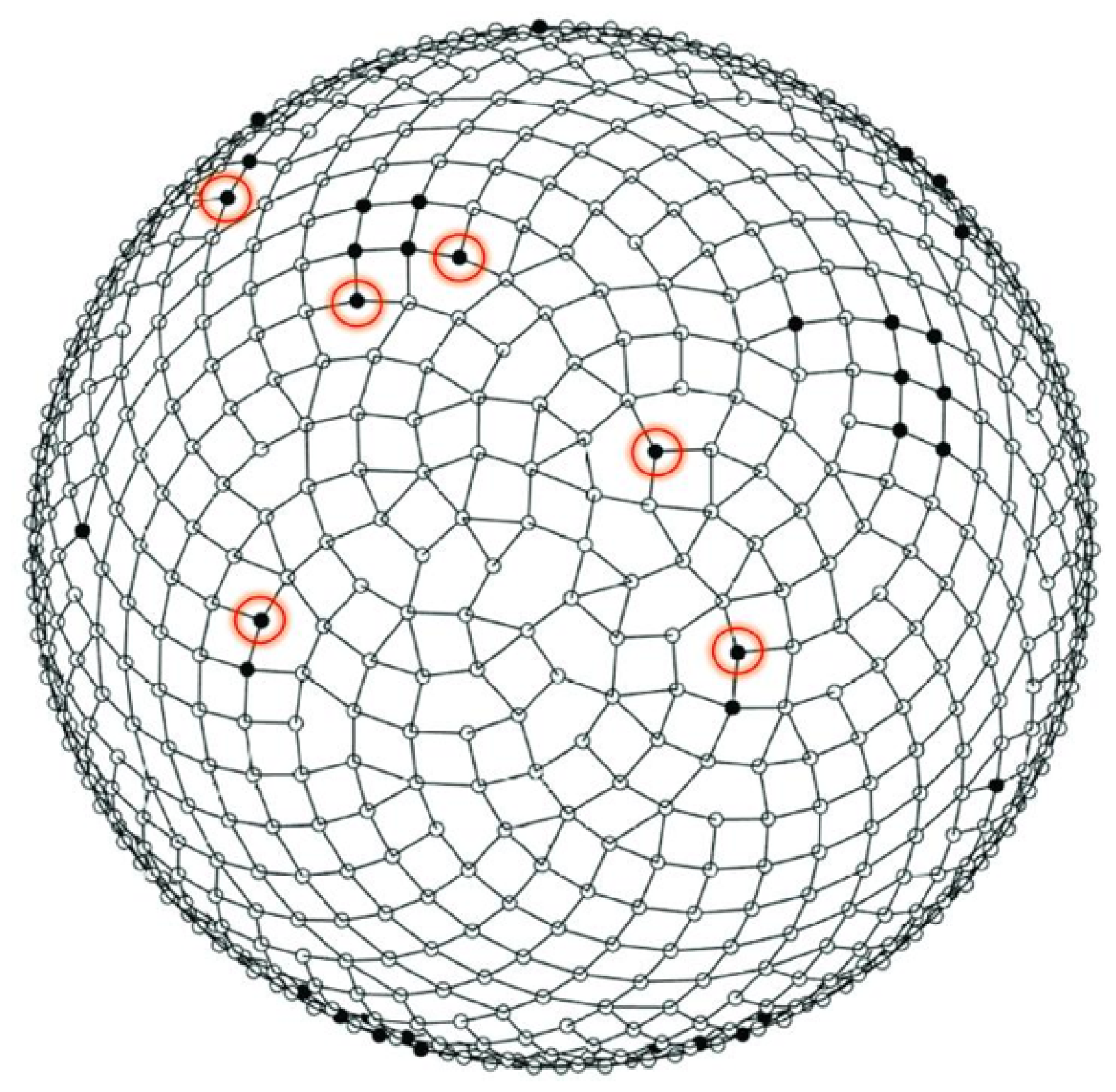}
\includegraphics[width=2.03in]{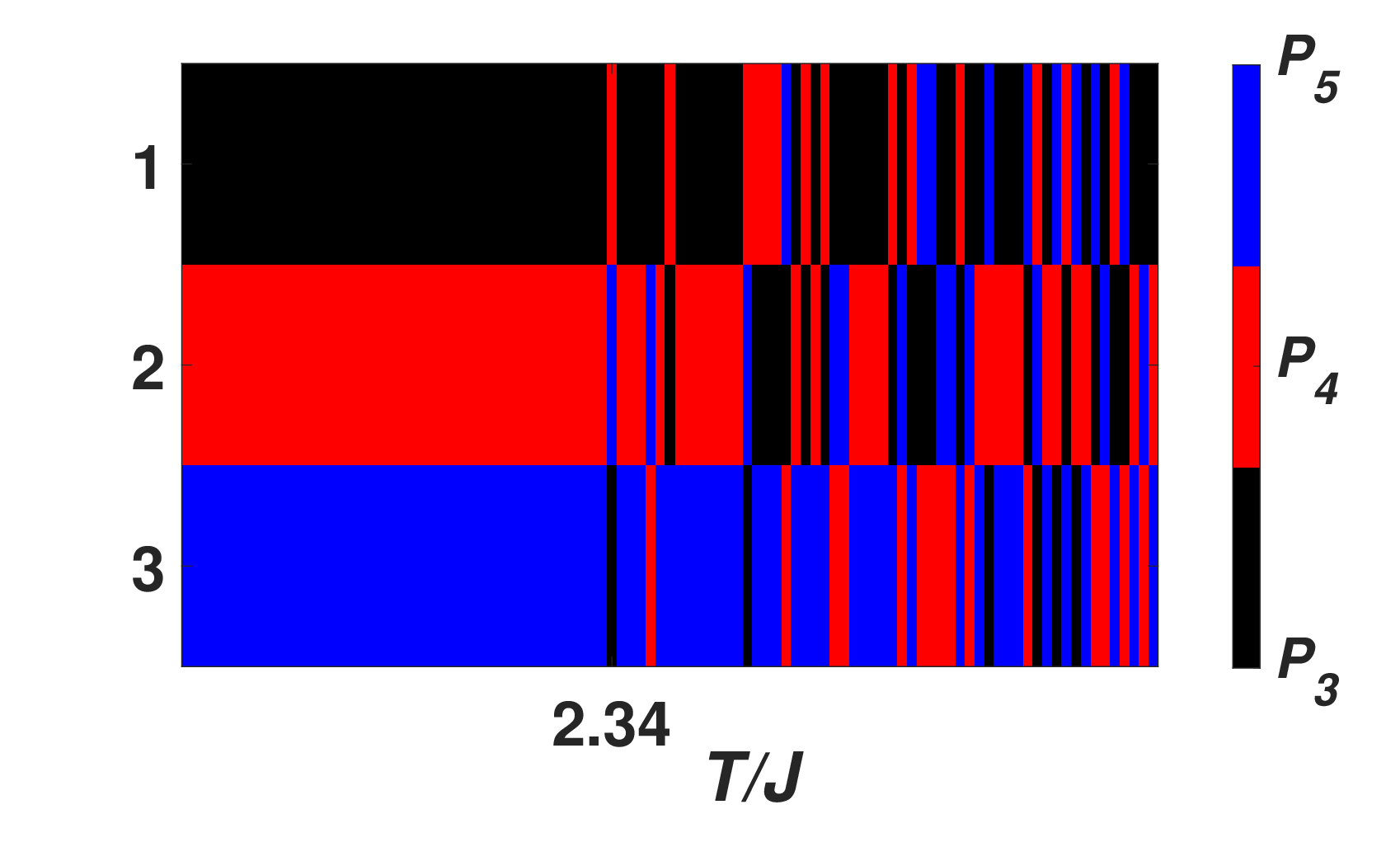}\\
\includegraphics[width=3.3in]{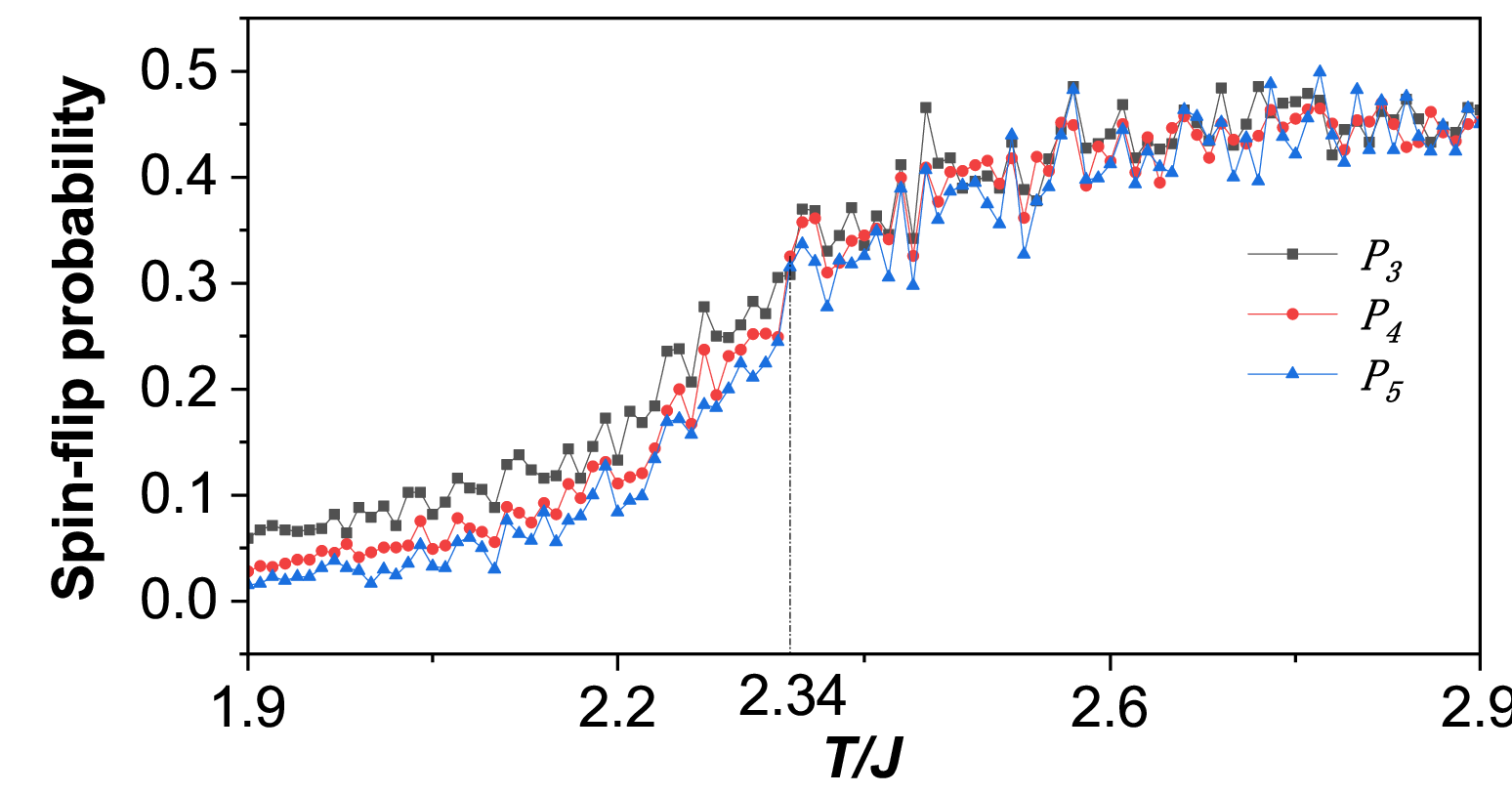}
\caption{(Top panel) A perspective of the  Fibonacci lattice at $T/J=2.0$, in which the spins at 6 sites with three neighbours change direction (Left). The ranking of the statistical spin-flip probabilities $P_3$, $P_4$ and $P_5$ at different temperatures (Right). (Bottom panel) Statistical spin-flip probabilities as functions of temperature. }	\label{Flippro}
\end{figure}

Note that the Fibonacci lattice is inhomogeneous, with varying probabilities of spin flips at different sites, which is crucial for understanding stable spin patterns. For instance, a site with three neighbors incurs an energy cost of $6J$ to flip its spin, while a site with five neighbors incurs a cost of $10J$. Consequently, the probability of a spin-flip is relatively higher for the site with three neighbors. 
In the left subpanel of the top row in Fig.\ref{Flippro}, we present a view of the Fibonacci lattice from a particular direction at $T/J=2.0$. It is evident that among the sites where the spin is flipped, at least 6 out of 76 sites have three neighbors. In contrast, only 11 out of 850 sites with four neighbors exhibit spin flips. This indicates that the proportion of spin flips is relatively significant for sites with three neighbors. We also statistically calculate the average spin-flip probabilities $P_3$, $P_4$ and $P_5$ for sites with 3, 4 and 5 neighbours, respectively, across different stable spin configurations at various temperatures. The results are illustrated in the bottom panel of Fig.\ref{Flippro}, with black squares representing $P_3$, red disks representing $P_4$, and blue triangles representing $P_5$.
To better illustrate the temperature-dependent changes in the relative spin-flip probabilities for different neighboring sites, we display the ranking of $P_3$, $P_4$ and $P_5$ as a function of temperature in the right subpanel of the top row in Fig.\ref{Flippro}, with black, red and blue squares representing $P_3$, $P_4$, and $P_5$ respectively. The numbers on the left, ranging from 1 to 3, indicate the relative ranking of the values rather than their specific numerical values.

Specifically, we perform 100 independent Monte Carlo simulations at each temperature to obtain 100 independent spin configurations. Each simulation is initialized with a random spin configuration and evolved according to the Metropolis criterion for $5\times 10^5$ steps to ensure thermal equilibration. In each simulation, we count the number of spin flips occurring at sites with 3, 4, and 5 neighbours in the stable spin configurations and divide these counts by the total number of sites of the corresponding type to obtain the flip ratios. Finally, the results from the 100 simulations at each temperature are averaged to obtain statistically reliable spin-flip probabilities.
At low temperatures, $P_3>P_4>P_5$, reflecting that sites with fewer neighbors are more prone to spin flips. As the temperature increases to  $T/J\gtrsim 2.34$, thermal fluctuations become more significant relative to the energy cost of spin flips, leading to a disruption in the ordering of $P_3$, $P_4$ and $P_5$. The bottom panel displays that $P_4>P_5>P_3$ at $T/J=2.34$. This effect has a significant impact on the phase transition temperature, as will be demonstrated later.

\begin{figure}[ht]
\centering
  \includegraphics[width=3.1in]{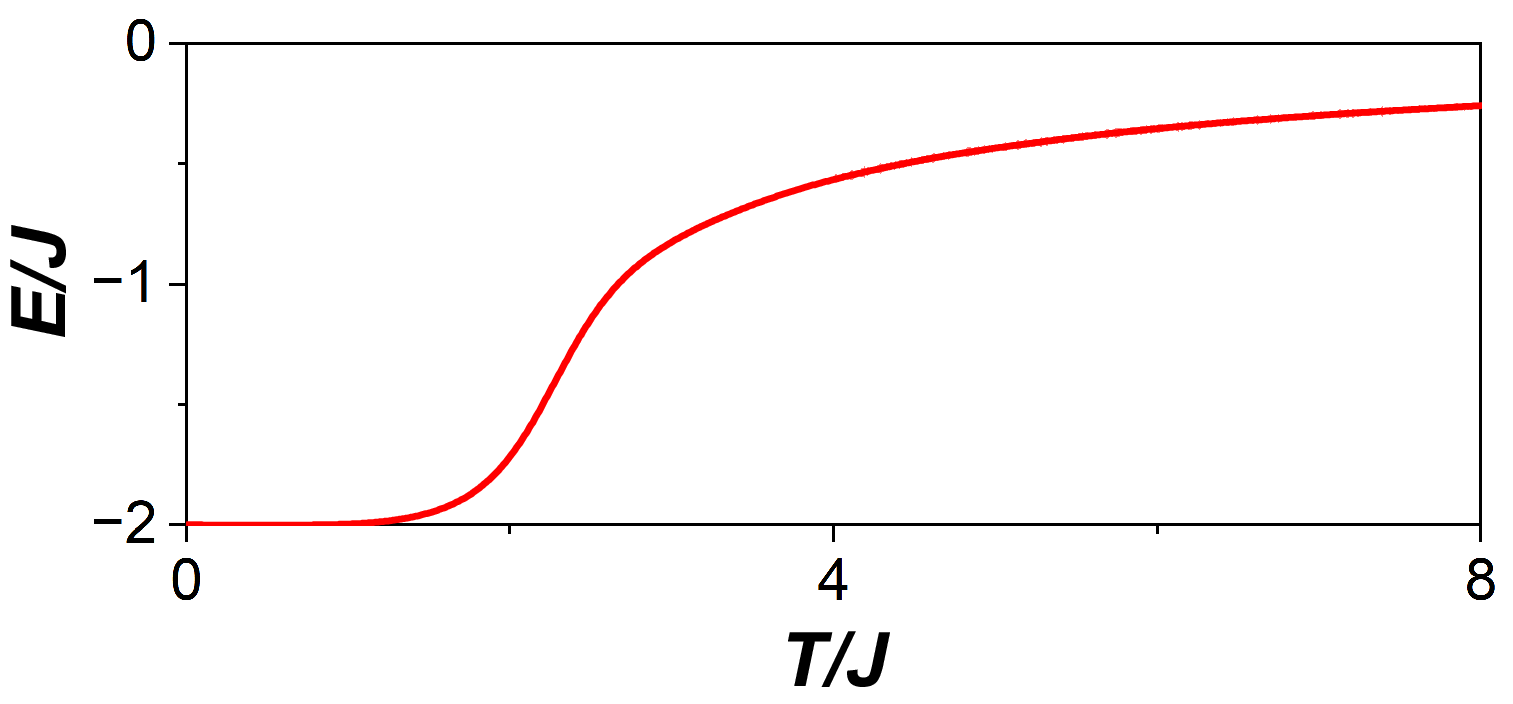} \\
  \includegraphics[width=3.1in]{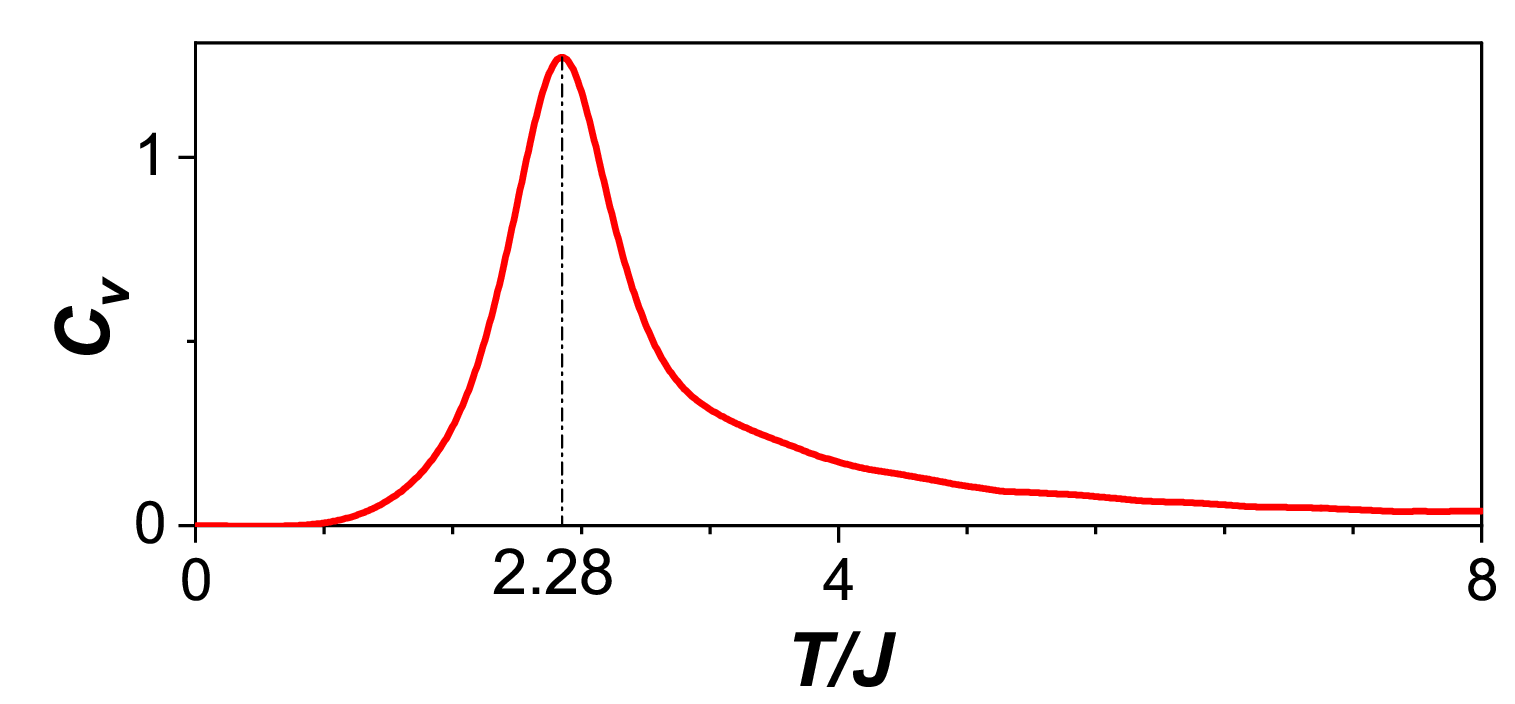}\\
  \includegraphics[width=3.1in]{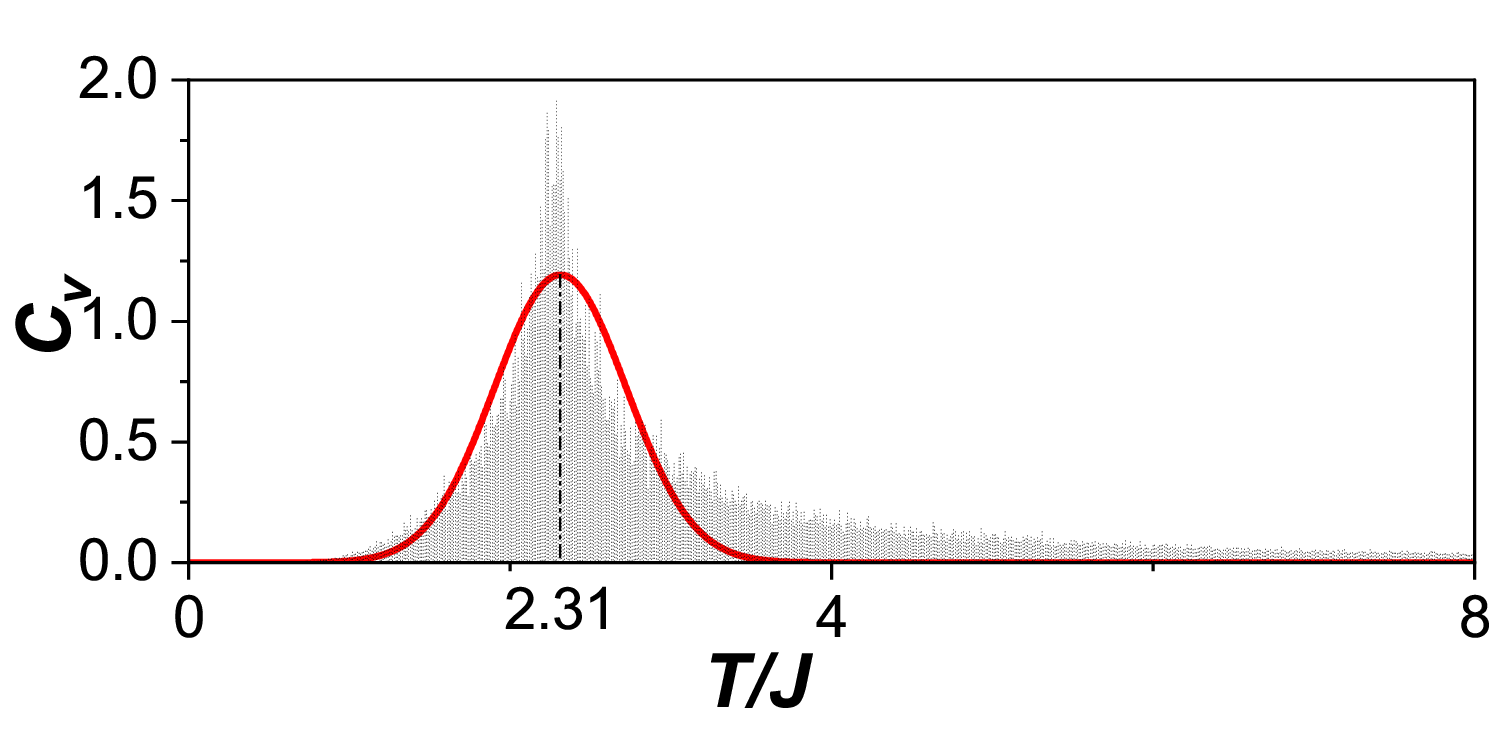}
  \caption{(Top panel) Total energy as a function of temperature for the spherical ferromagnetic Ising model with $N=1000$. (Middle panel) Specific heat as a function of temperature, calculated using $C_V=\left(\frac{\partial E}{\partial T}\right)_V$.
(Bottom panel) Specific heat as a function of temperature, calculated using Eq.(\ref{Cveqn}). The black dashed lines represent the raw Monte Carlo data points, and the red solid line shows the Gaussian fit around the peak.
}	\label{Cv1}
 \end{figure}

\subsection{Determining $T_c$ via the behaviors of the specific heat and the magnetic susceptibility}

Next, we aim to determine the phase transition temperature $T_c$ of the spherical Fibonacci Ising model. To achieve this, we employ three different approaches. The first one is straightforward but relatively approximate: We estimate $T_c$ by analyzing the behavior of the specific heat $C_V$ using three different methods for cross-validation.
The second one involves estimating $T_c$ by analyzing the magnetization $M$ and magnetic susceptibility $\chi$.
The third one offers greater accuracy and involves using graph convolutional networks (GCNs) to refine our estimation of $T_c$.

Our first method to estimate the specific heat utilizes the formula $C_V=\left(\frac{\partial E}{\partial T}\right)_V$. 
Given a stable spin configuration, the total energy $E$ of the system is calculated via Eq.(\ref{e1}). We then average the total energy over 100 stable spin configuration at each temperature, with a temperature interval $0.01J$. The numerical results are presented in the top panel of Fig.\ref{Cv1}. To estimate the specific heat, we use the spline functions to fit the $E$ versus $T$ curve, resulting in  a function $E(T)$. We subsequently compute the derivative of this function to estimate $C_V$ and plot $C_V$ vs $T$ in the middle panel of Fig.\ref{Cv1}. As depicted in the figure, $C_V$ exhibits a peak behavior at $T/J\approx 2.28$, which can be considered as an estimate of $T_c$. Interestingly, this value is very close to the critical temperature of the two dimensional Ising model on a square lattice, which is $T^\Box_c/J= 2.269$ \cite{PhysRev.60.252,PhysRev.65.117,PhysRev.76.1232}.  To quantify the uncertainty in this estimation of $T_c$, we apply a nonparametric Bootstrap resampling approach \cite{efron1992bootstrap,kawano1995bootstrap}, which has recently been applied to the Ising model \cite{LCYThesis}. Further details of the method are provided in Appendix \ref{app1}. The bootstrap iteration is performed $B=1000$ times, yielding a more precise estimate of $T_c$, namely $T_c/J=2.279 \pm 0.005$.

The second method for estimating $C_V$ is based on the the fluctuation-dissipation formula for specific heat:
\begin{equation}\label{Cveqn}
C_V =\frac{1}{T^2}\frac{\langle E^2 \rangle - \langle E \rangle^2}{N},
\end{equation}
where the Boltzmann constant is set to 1 in natural units. Here, the energy $E$ is obtained using the previous method, and its absolute value is quite large. Therefore, when calculating $C_V$ by using Eq.(\ref{Cveqn}), the squared terms become even larger, which can significantly amplify the oscillations of $E$. 
However, the results from the previous method may be influenced by the curve fitting technique. In contrast, this method does not require calculating the slope of the fitted curve, thereby avoiding errors caused by different fitting approaches.
In the bottom panel of Fig.\ref{Cv1}, we present our calculations based on Eq.(\ref{Cveqn}), where each data point is averaged over 100 stable spin patterns too.
As expected, the value of $C_V$ exhibits significant oscillations as the temperature changes. However, there is also a distinct maximum around $T/J \approx 2.313$. We further attempt to fit the curve using a Gaussian function and observe that the peak is indeed located around $2.313J$. To improve the accuracy of our results, we apply the Bootstrap error analysis, which refines the estimation of $T_c/J$ to $2.31 \pm 0.007$. Although this method introduces a larger uncertainty, the confidence interval of the results still overlaps with that obtained from the previous method.

\begin{figure}[ht]
\centering
  \includegraphics[width=3.1in]{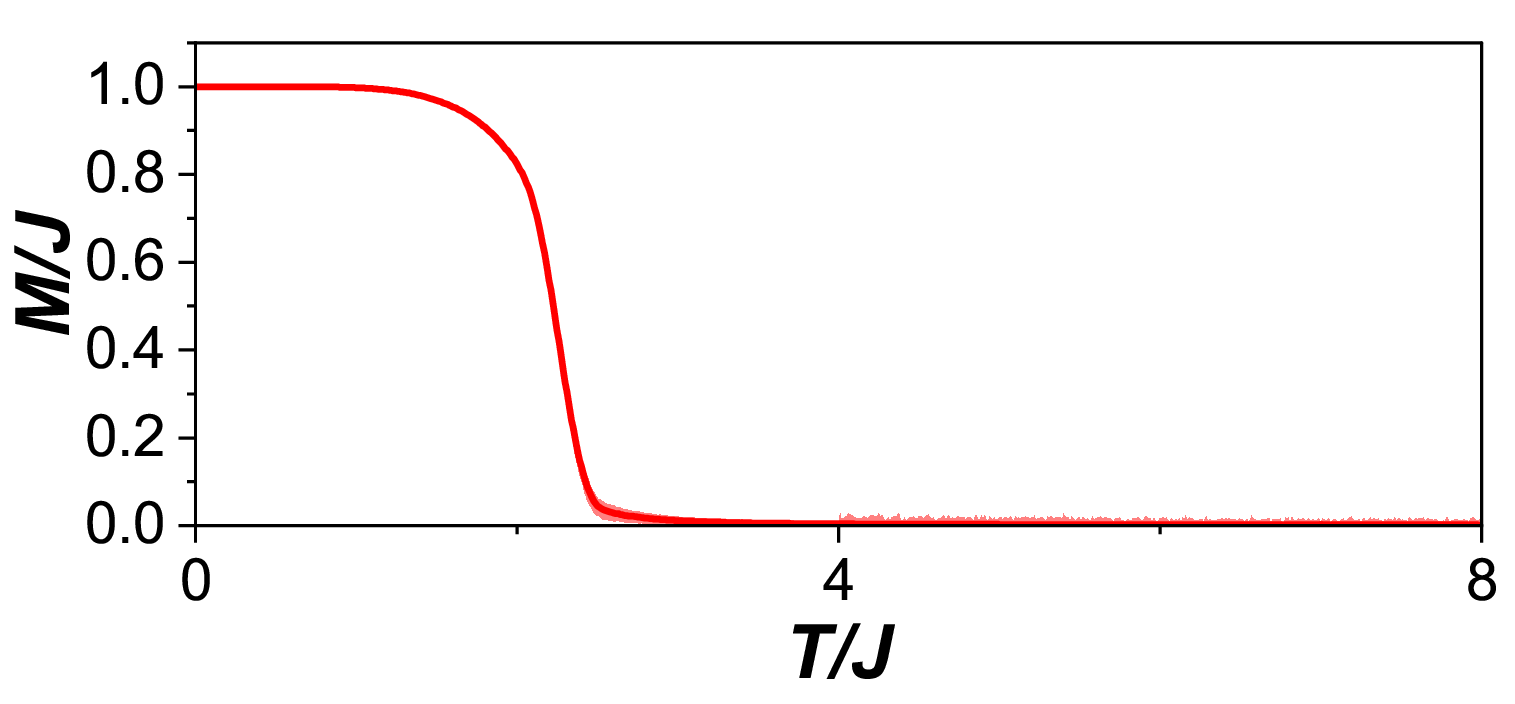} \\
  \includegraphics[width=3.1in]{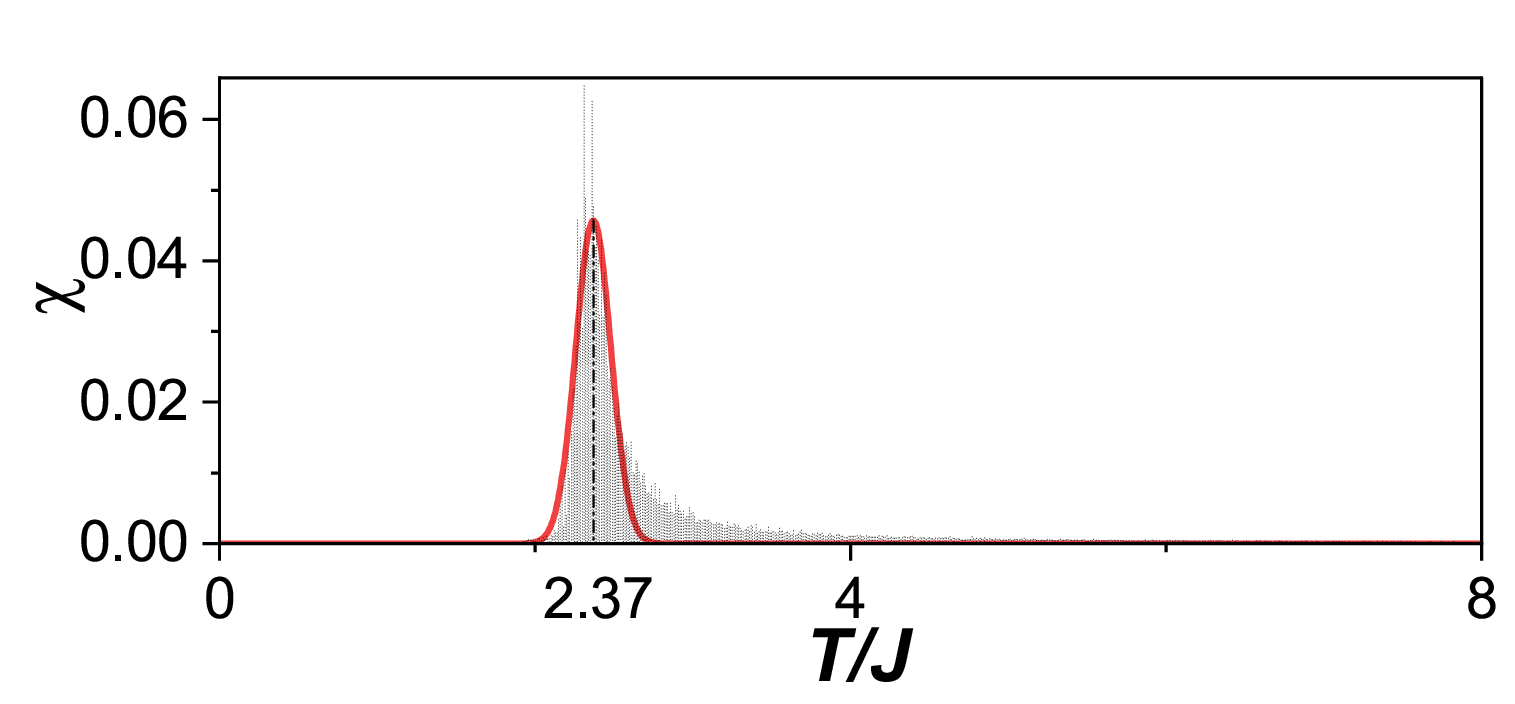}
  \caption{Top panel) Magnetization as a function of temperature for the spherical ferromagnetic Ising model with $N=1000$. (Bottom panel) Magnetic susceptibility as a function of temperature, calculated using $\chi = \frac{\langle M^2 \rangle - \langle M \rangle^2}{NT}$. The black dashed lines represent the raw Monte Carlo data points, and the red solid line shows the Gaussian fit around the peak.}	\label{MS1}
 \end{figure}
The phase transition temperature $T_c$ can also be estimated by analyzing the variation of the magnetization $M$ and magnetic susceptibility $\chi$ with temperature.
Given a stable spin configuration, the magnetization $M$ and susceptibility $\chi$ of the system can be calculated by
\begin{equation}\label{MS}
M=\sum_{i} s_{i}, \qquad \chi = \frac{\langle M^2 \rangle - \langle M \rangle^2}{NT}.
\end{equation}
Through methods similar to those used for analyzing the temperature dependence of energy and specific heat, we average the total magnetization and susceptibility over 100 stable spin configuration at each temperature, with a temperature interval $0.01J$. The numerical results are presented in Fig.\ref{MS1}, where the top panel shows the variation of the magnetization with temperature, and the bottom panel shows the variation of the susceptibility with temperature. As depicted in the figure, the susceptibility $\chi$ exhibits a peak behavior at $T/J\approx 2.369$, which can be considered as an estimate of $T_c$. We also apply the Bootstrap error analysis to improve the accuracy of the results, which refines the estimation of $T_c/J$ to $2.370 \pm 0.005$, which is close to the estimates of $T_c$ obtained from the specific heat analysis.

\subsection{Determining $T_c$ via the method of GCNs}

The second approach employs machine learning, which has proven to be a powerful tool in processing big data \cite{MLPRB16,MLS17,MLNP17,MLPRX17,MLPRX17b,MLPRE17,MLPRL18a,MLPRL18b}. Given that the Fibonacci lattice is not homogeneous and does not resemble a typical image-like structure, traditional CNNs can not be applied to this model \cite{CNNPRB18,CNNPRE19}. Instead, we utilize GCNs, which are capable of capturing spatial structural features by leveraging the connectivity relationships between nodes \cite{GCN20}. This makes GCNs particularly well-suited for processing topologically structured data like the Fibonacci lattice. The GCNs does not require explicit geometric coordinates of the spherical Fibonacci lattice (e.g., site latitude and longitude) and can capture local spin patterns through the adjacency matrix based on a nearest-neighbor cutoff radius 
$r_s$. Because strong correlations between spins imprint phase transition information onto local patterns, the GCNs can infer global phases from local structures. This behavior was also observed in our previous study of the spherical XY model \cite{PhysRevResearch.4.023005}. Details of GCNs implementation can be found in Appendix \ref{app3}.

\begin{figure}[ht]
\centering
  \includegraphics[width=3.1in]{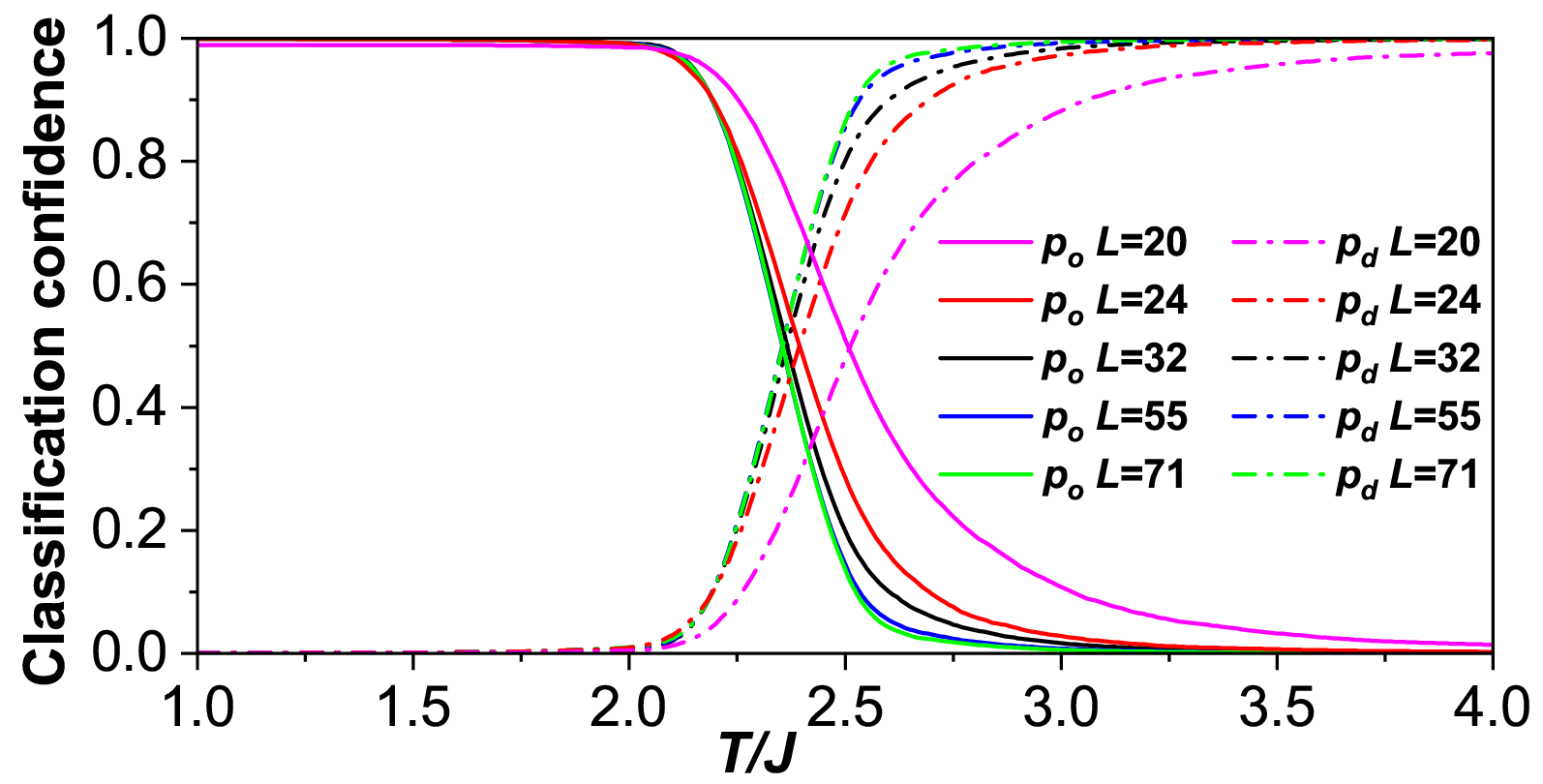}
  \caption{Classification confidences $p_\text{o}$ and $p_\text{d}$ as a function of temperature for the 2D Ising model on a $L\times L$ square lattice with periodic boundary conditions. The pink, red, black, blue, and green lines correspond to lattice sizes $L=20, 24, 32, 55,$ and $71$, respectively.}	\label{2D}
 \end{figure}

To determine the critical temperature $ T_c $ using GCNs, we adopt a supervised learning approach inspired by prior studies [33], but tailored to our system. Unlike traditional methods that rely on training samples labeled by a classifier with prior knowledge of $T_c$, we take advantage of the clear separation of phases at extreme temperatures to construct our training set. Specifically, MC spin configurations are generated at very low temperatures (e.g., $ T/J = 0.01 $, well below the expected $ T_c $) and very high temperatures (e.g., $ T/J = 8.0 $, well above the expected $ T_c $), where the system is unambiguously in the ordered and disordered phases, respectively. These configurations are labeled as ``ordered" ($ p_\text{o} = 1$, $p_\text{d} = 0 $) and ``disordered" ($ p_\text{o} = 0$, $p_\text{d} = 1 $) based on their temperature regimes, eliminating the need for an exact $ T_c $ value a priori.
Here $ p_\text{o}$ and $p_\text{d}$ represent the classification confidences for the ordered and disordered phases, respectively.
The trained GCNs then interpolates across the temperature range $[0.01, 8.0]J$, predicting classification confidences $ p_\text{o} $ and $ p_\text{d} $ for each temperature. The intersection point where $ p_\text{o} = p_\text{d} $ is taken as the estimated $ T_c $.

\begin{figure}[ht]
\centering
  \includegraphics[width=3.1in]{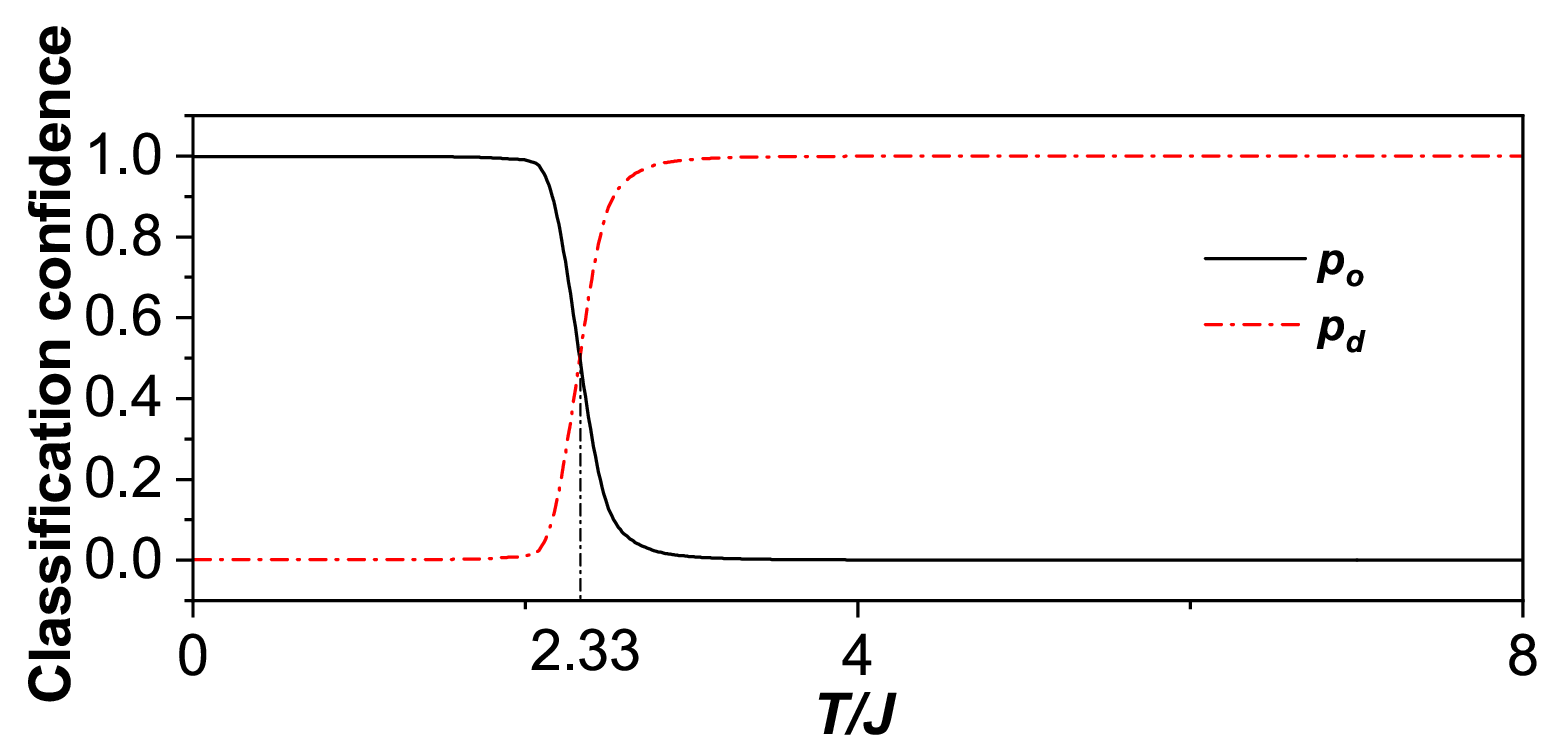} \\
   \includegraphics[width=3.1in]{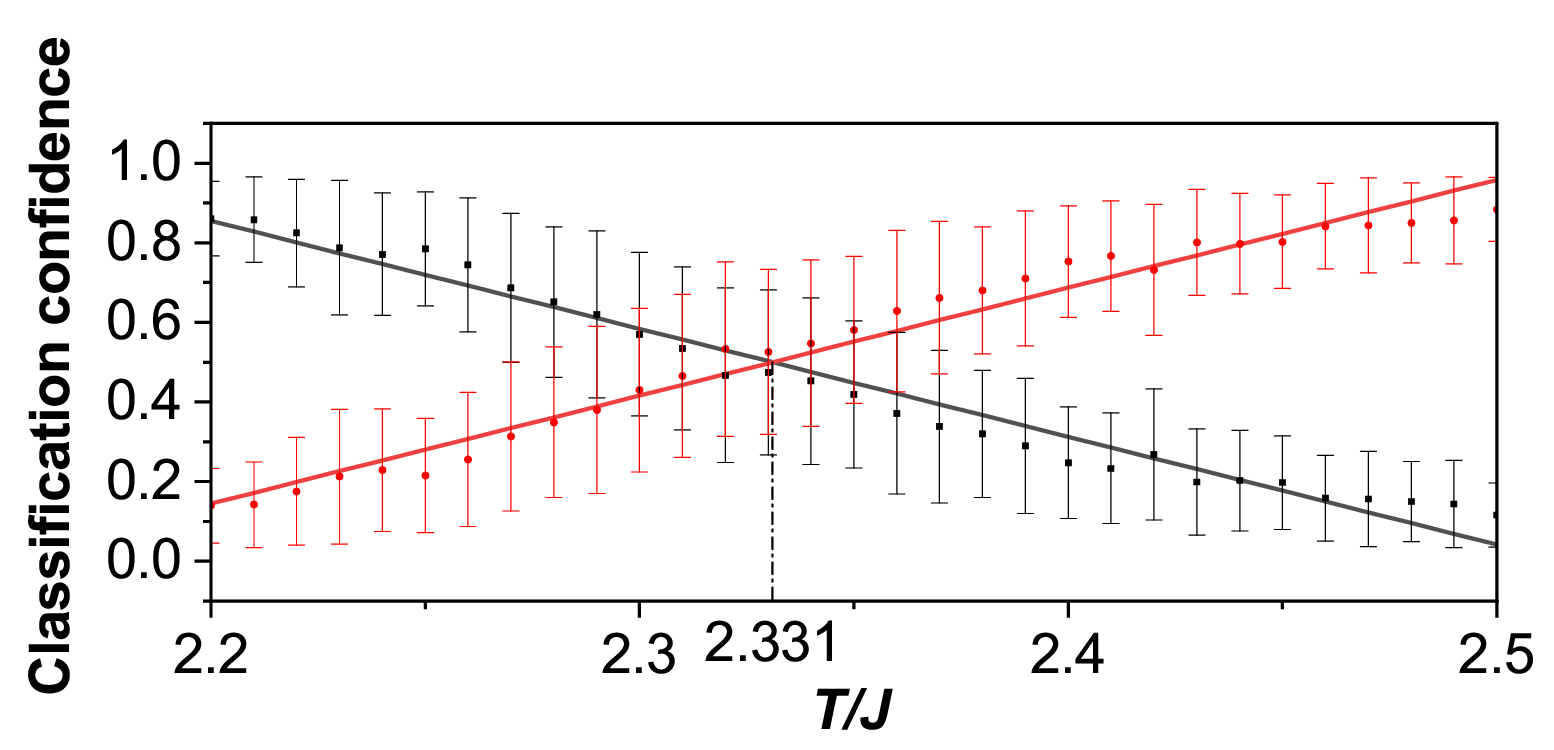}
  \caption{(Top Panel) Classification confidences, $p_\text{o}$ and $p_\text{d}$, versus temperature for the ferromagnetic Ising model on a spherical Fibonacci lattice. The phase transition temperature $T_{c}$ is the point at which $p_\text{o}=p_\text{d}$.
  (Bottom panel) Error analysis of $T_c$ via linear fitting near the point $p_\text{o}=p_\text{d}$.}\label{Tc1}
 \end{figure}

To verify the accuracy of the GCNs simulations, we first conducted preliminary tests on two-dimensional square lattices with periodic boundary conditions, with lattice sizes of $20 \times 20$, $24 \times 24$, $32 \times 32$, $55 \times 55$, and $71 \times 71$, in order to compare with the spherical Ising model containing 500, 1000, 3000, and 5000 sites discussed later. As shown in Fig. \ref{2D}, the solid lines represent the classification confidence for the ordered phase, while the dash-dot lines represent the disordered phase. As the temperature increases, the solid lines gradually decrease from 1, indicating a reduced similarity between the simulation results and the ordered phase, whereas the dash-dot lines steadily increase from 0, reflecting an increasing similarity with the disordered phase. The phase transition temperatures for different lattice sizes $L = 20, 24, 32, 55, 71$ can be determined from the temperatures at which the classification confidences $p_\text{o}$ and $p_\text{d}$ intersect, yielding $T_c/J = 2.51, 2.40, 2.36, 2.35, 2.35$, respectively.
 Through the simulation and analyses of different lattice scales, it is found that as the number of lattice sites increases, the phase transition temperature gradually approaches the theoretical value of $T_c/J = 2.269 $. Specifically, the values obtained are $T_c/J = 2.51 $ for the $20 \times 20 $ lattice and $T_c/J = 2.35 $ for the $71 \times 71 $ lattice. The latter demonstrates strong concordance with the predicted critical temperature. This indicates that increasing the lattice size can enhance the accuracy of phase transition temperature estimations. Moreover, the GCNs effectively captures the phase transition behavior and critical properties.

After validating the accuracy of GCNs, we subsequently apply it to determine the phase transition temperature of the spherical Ising model. In numerical calculations, we select 800 temperature nodes within the range of $[0.01, 8.0]J$, with an interval of $0.01J$, and perform 100 simulations for each temperature node. The results, presented in the top panel of Fig.\ref{Tc1}, indicate that the critical temperature is determined to be $T_c/J=2.33$, where  $p_\text{o}=p_\text{d}$. This value is fairly close to those obtained from the analysis of the specific heat $C_V$, but it is comparatively
higher than that of the square lattice, $T^\Box_c= 2.269J$. 
Due to the inhomogeneity of the Fibonacci lattice, a complete analytical analysis is not feasible. We believe a possible reason for this is the presence of sites with 3 or 5 neighbors, which affects the spin dynamics within the system.
At low temperatures, a site with a greater number of neighbors is less likely to experience a spin flip, as confirmed in Fig.\ref{Flippro}. However, as the temperature increases beyond $T/J \gtrsim2.34$, the situation changes. Fig.\ref{Flippro} shows that at $T/J=2.34$, the spin-flip probabilities follow the order $P_4>P_5>P_3$ at $T/J=2.34$, indicating that spins on sites with 3 or 5 neighbors are more resistant to flipping. This, to some extent, enhances the ``stability" of the ordered phase, effectively elevating the phase transition temperature.

To analyze the uncertainty of $ T_c $, we note that it is determined by the intersection of the two curves $ p_\text{o}(T) $ and $  p_\text{d}(T) $, both of which exhibit good linearity near $ T_c $. Therefore, the uncertainty can be estimated as follows.
We select an interval $[T_1, T_2]$ around $ T_c $, where $  p_\text{o}(T) $ and $  p_\text{d}(T) $ can be approximated as linear functions with slopes $ k_\text{o} $ and $ k_\text{d} $, respectively. Each point on these two curves has its own uncertainty, obtained by averaging over 100 stable spin configurations. Using error propagation, we can determine the uncertainties of $ k_\text{o} $ and $ k_\text{d} $, denoted as $ \sigma_\text{o} $ and $ \sigma_\text{d} $, respectively. Finally, the uncertainty of $ T_c $ is given by
\begin{align}\label{kerror}
\Delta T_c=\frac{\sqrt{\sigma_\text{o}^2+\sigma_\text{d}^2}}{|k_\text{o}-k_\text{d}|}.
\end{align}
In the bottom panel of Fig.\ref{Tc1}, we present our numerical results, yielding $T_c/J = 2.331 \pm 0.090$. Note the value $2.331J$ is obtained from the intersection of the two linear fits, which differs slightly from $T_c/J=2.33$, determined by the condition $p_\text{o}=p_\text{d}$.

While the GCNs machine learning method is not intended to replace traditional Monte Carlo simulations, it serves as a complementary tool to efficiently extract patterns and identify phase transitions from large configuration datasets. GCNs is particularly advantageous for automatically capturing local spin patterns on non-uniform lattices, quickly estimating critical temperatures via classification confidence and decision boundaries, and adapting to complex geometries and multi-phase systems.

\subsection{Phase Transition Detection via PCA and KMeans Clustering)}
\begin{figure}[ht]
\centering
  \includegraphics[width=1.5in]{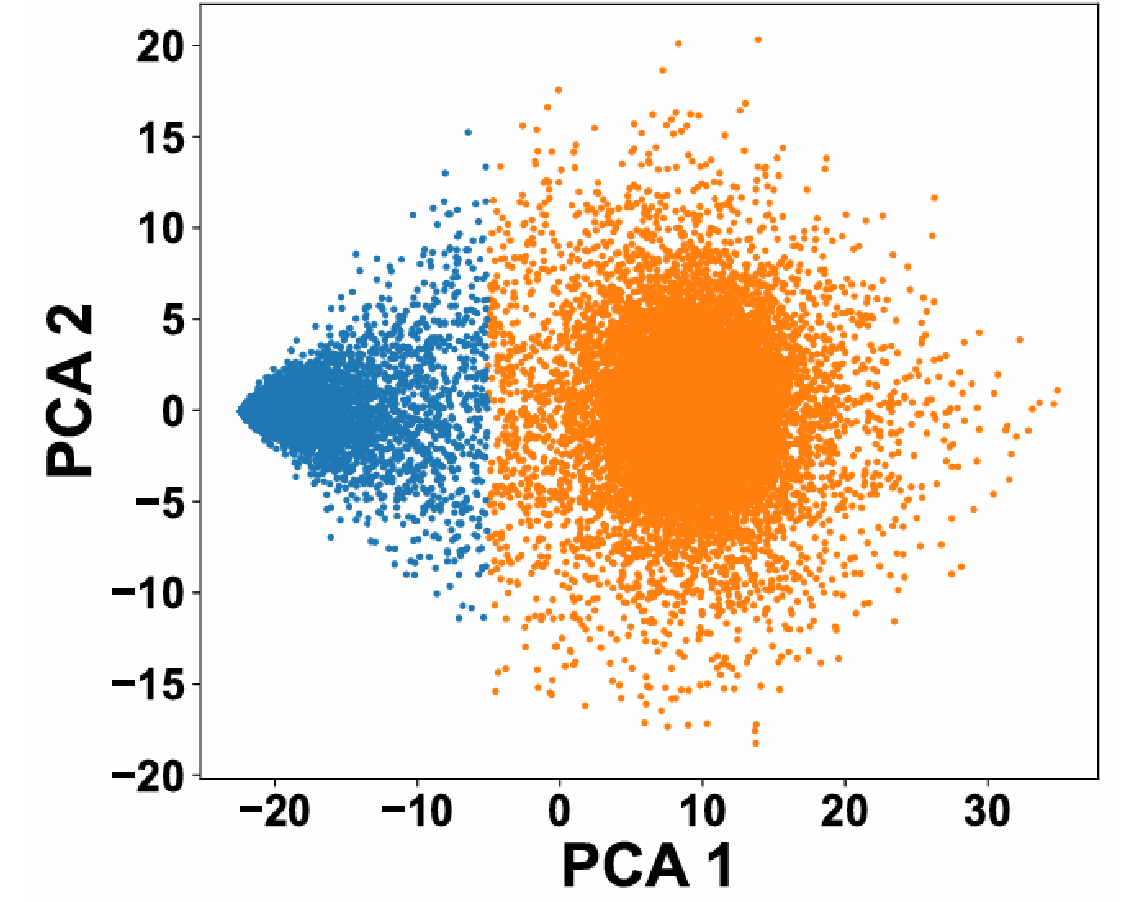}
  \includegraphics[width=1.5in]{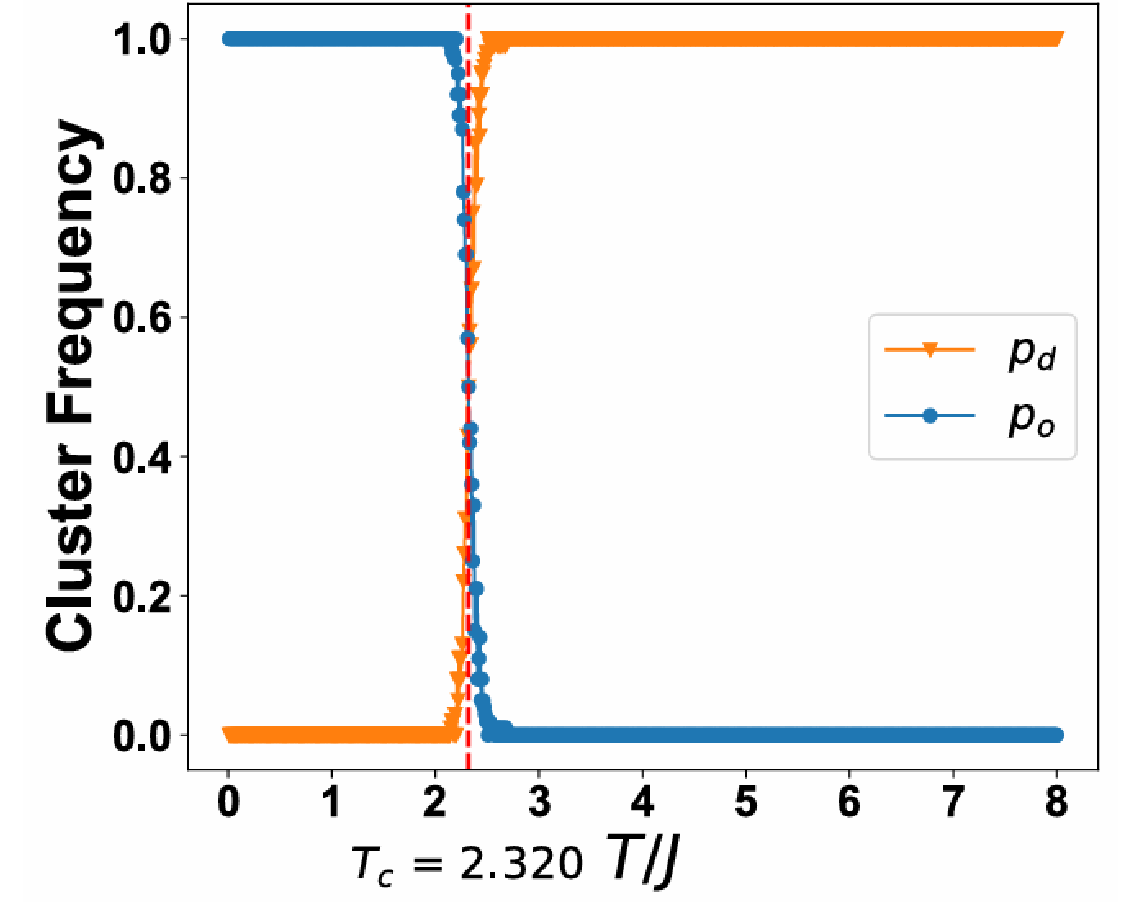} 
  \caption{(Left panel) The similarity of different spin configurations with respect to the two principal components. (Right panel)The cluster fraction, $p_\text{o}$ and $p_\text{d}$, versus temperature of the ferromagnetic Ising model on a spherical Fibonacci lattice for $N=1000$.
  }	\label{PCA}
 \end{figure}
 
To further validate the phase transition temperatures in the spherical ferromagnetic Ising model for $N=1000$, we employed an unsupervised learning method combining Principal Component Analysis (PCA) and KMeans clustering \cite{abdi2010principal, lancia2013equilibrium, wetzel2017unsupervised}. This approach does not rely on predefined order parameters, making it well-suited for detecting statistical structure changes in complex spin systems.

We first applied PCA to the spin configurations sampled at various temperatures, each represented as a 1000-dimensional vector of ±1 values. The dimensionality was reduced to two principal components, and the resulting data were clustered using KMeans (with number of clusters set to 2). For each temperature, we calculated the fraction of samples that fall into each cluster, defined as the cluster fraction, and plotted the cluster fraction as a function of temperature.

As shown in the left panel of Fig.\ref{PCA}, the samples are clearly grouped into two clusters, indicating that this method can effectively distinguish between the ordered and disordered phases. The right panel shows a sharp change in cluster fractions as the temperature increases, indicating a phase transition between two distinct phases. The transition point corresponds to the system’s critical temperature, estimated to be around $T_c/J\approx 2.32$. This result aligns with the peak of the specific heat and the critical temperature predicted by GCNs-based classification.

This unsupervised PCA plus clustering pipeline offers a powerful and intuitive visualization tool for phase transition detection, particularly in systems with curved geometry or non-uniform lattices where conventional order parameters are less accessible.

\subsection{Effects of $r_c$ and $N$}

\begin{figure}[ht]
\centering
  \includegraphics[width=3.1in]{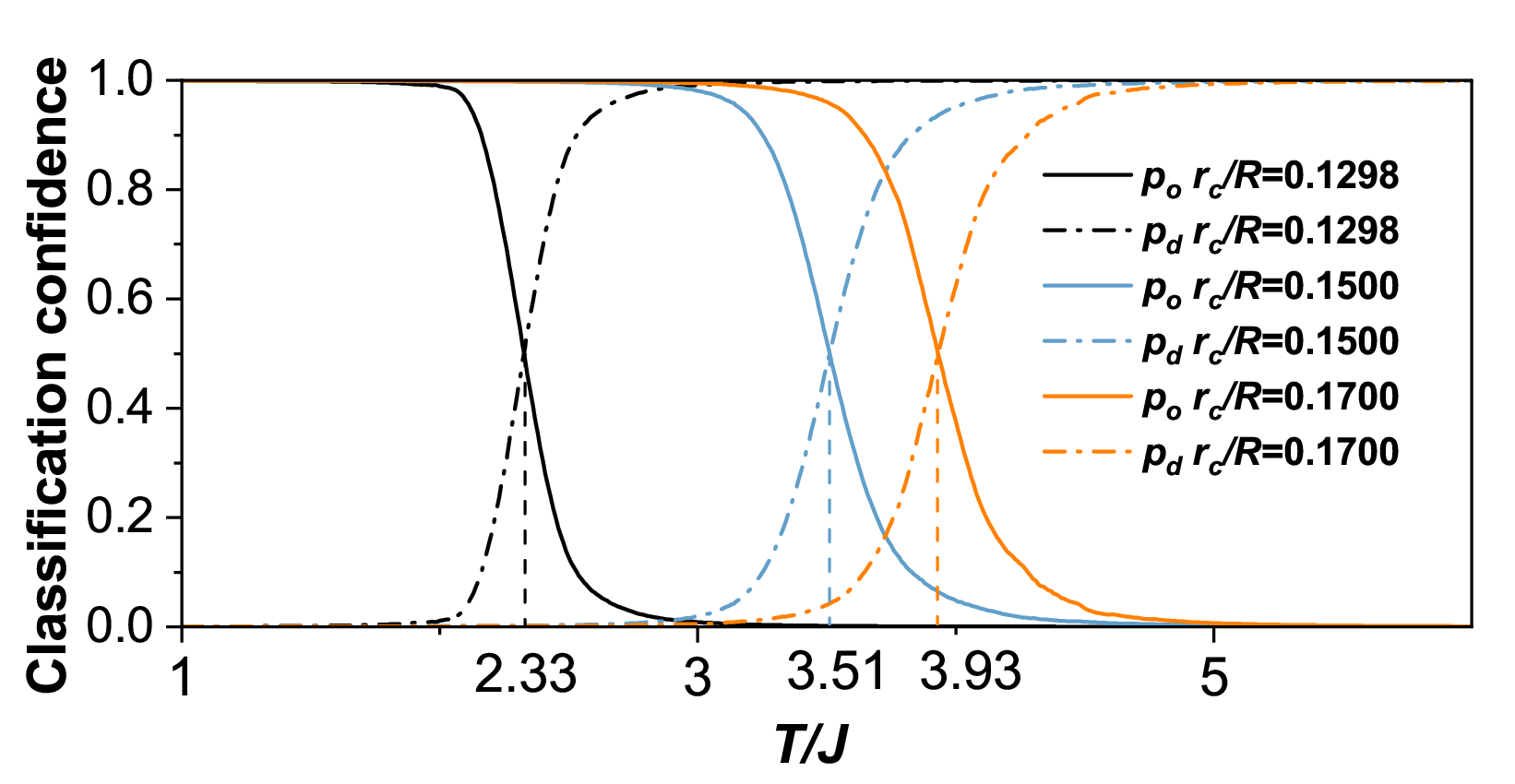} \\
    \includegraphics[width=3.1in]{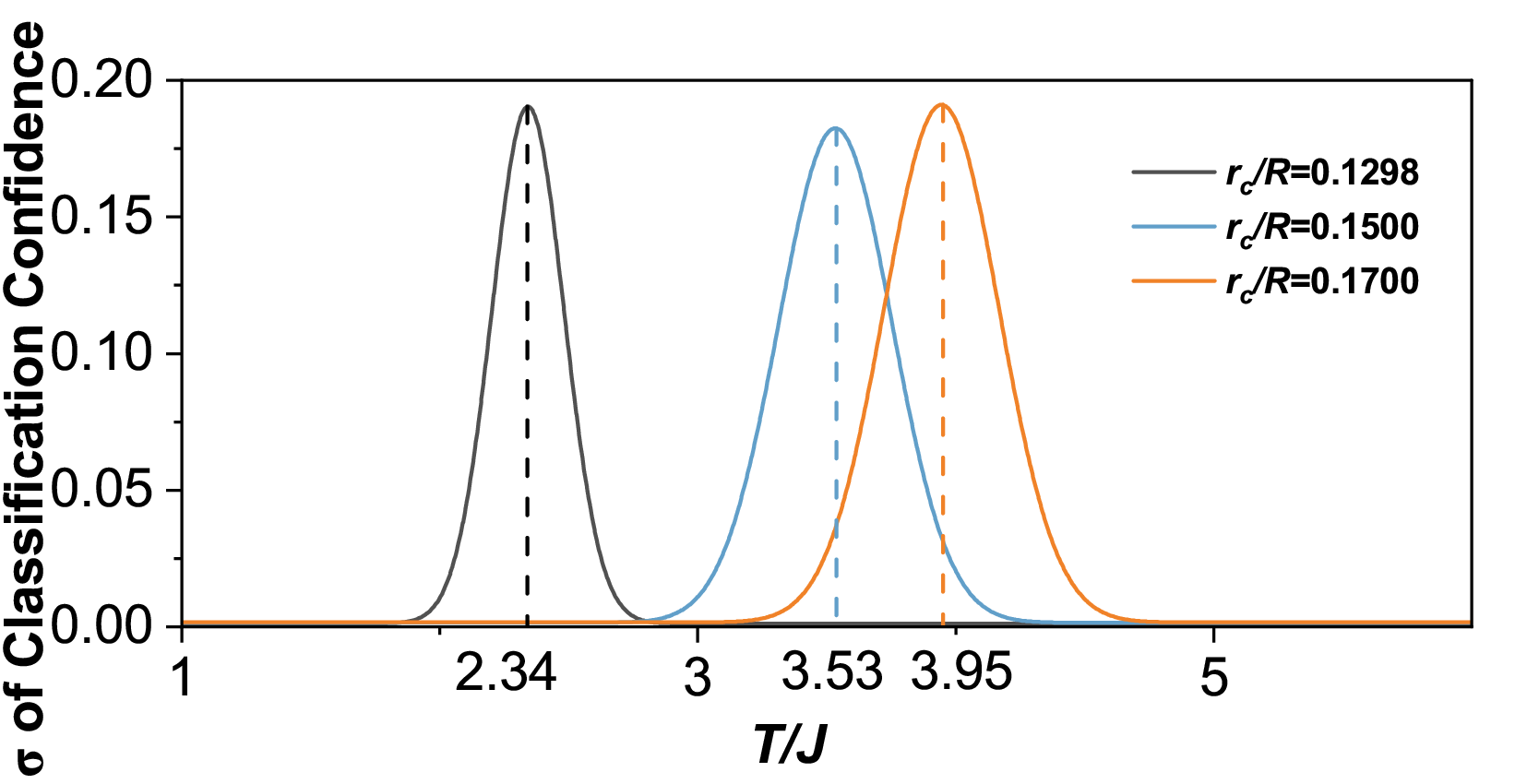}
  \begin{tabular}{|c|c|c|c|c|c|c|}
\hline
 \textbf{$r_c/R$} & \makecell{$N_3$\\ ($w_3$)} & \makecell{$N_4$\\ ($w_4$)} & \makecell{$N_5$\\ ($w_5$)}& \makecell{$N_6$\\ ($w_6$)} & \makecell{$N_7$\\ ($w_7$)} & \makecell{$N_8$\\ ($w_8$)} \\ \hline
0.1298 & \makecell{76 \\(7.6\%) }& \makecell{850\\ (85\%) }& \makecell{74\\ (7.4\%)} & - & - & - \\ \hline
  0.1500 & - & \makecell{26 \\(2.6\%)} & \makecell{240 \\(24\%)} & \makecell{734 \\(73.4\%) }& - & - \\ \hline
 0.1700 & - & - & \makecell{8 \\(0.8\%)} & \makecell{722 \\(72.2\%)} & \makecell{224 \\(22.4\%)} & \makecell{46\\ (4.6\%) }\\ \hline
\end{tabular}
  \caption{(Top panel) Classification confidences, $p_\text{o}$ and $p_\text{d}$, versus temperature for $N=1000$ and different values of $r_c$. 
  (Middle panel) Variances of the classification confidences as a function of temperature. 
  (Bottom panel) Percentage of sites with different neighbors for different  values of $r_c$. Here $N_i$ is the number of sites with $i$ neighbors, and $w_i=N_i/1000$.}	\label{MTc1}
 \end{figure}

Now, we further investigate the effects of other parameters, such as $r_c$ and $N$, on $T_c$ using Monte Carlo simulations and GCNs. First, we fix the total number of sites at
$N=1000$ and vary the interaction length $r_c$. As $r_c$ increases, each spin interacts with a greater number of spins, effectively increasing the number of neighbors per site and altering the lattice structure. In this scenario, as the temperature rises, the energy cost for a spin flip becomes larger due to the higher average number of neighbors per site, leading to an effective increase in $T_c$. We present our numerical results in Fig.\ref{MTc1}. The top panel shows that $T_c/J=2.33$ for $r_c/R=0.1298$, $T_c/J=3.51$ for $r_c/R=0.1500$, and $T_c/J=3.93$ for $r_c/R=0.1700$, which confirms our former reasoning. 
Moreover, the error made by GCNs when performing phase classification can serve as an indicator for the critical behavior of a dynamical system. Consequently, the temperature at which the classification error is maximized can itself be used as a phase transition criterion \cite{PhysRevD.102.054501}. To quantify this, we perform 100 independent simulations at each temperature point and compute the variance of the classification confidence. As shown in the middle panel, the maximum classification errors occur at 
$T_c/J=2.34$ for $r_c/R=0.1298$, $T_c/J=3.53$ for $r_c/R=0.1500$, and $T_c/J=3.95$ for $r_c/R=0.1700$, which agree well with the phase transition temperature obtained from the intersection of the two phases ($p_\text{o}$, $p_\text{d}$) in classification. This further supports the reliability of the maximum confusion criterion.
The bottom panel lists the fraction of sites with different numbers of neighbors for various $r_c$ values. For example, when $r_c=0.1500R$, $73.4\%$ of the sites have six neighbours, indicating that most of the lattice resembles a triangular structure. Interestingly, the estimated $T_c/J=3.51$ is relatively close to the exact value $T_c/J=4/\text{ln} 3\approx 3.64$ of the planar triangular lattice as reported in Ref.\cite{baxter2016exactly, PhysRev.79.357, PhysRevB.7.5017,HOUTAPPEL1950391,HOUTAPPEL1950425}. 
Since this model also includes some four-neighbor and five-neighbor sites, its spin flip energy is lower than that of the six-neighbor sites, which results in a relatively lower $T_c$. For $r_c/R=0.1700$,
 although a similar fraction of sites have six neighbors ($w_6=72.2\%$), the number of sites with more neighbors increases significantly ($w_7=22.4\%$ and $w_8=4.6\%$). This enhanced ``neighbor interaction" effectively strengthens spin correlations, leading to a higher phase transition temperature of $T_c/J=3.93$.

\begin{figure}[ht]
\centering
  \includegraphics[width=3.1in]{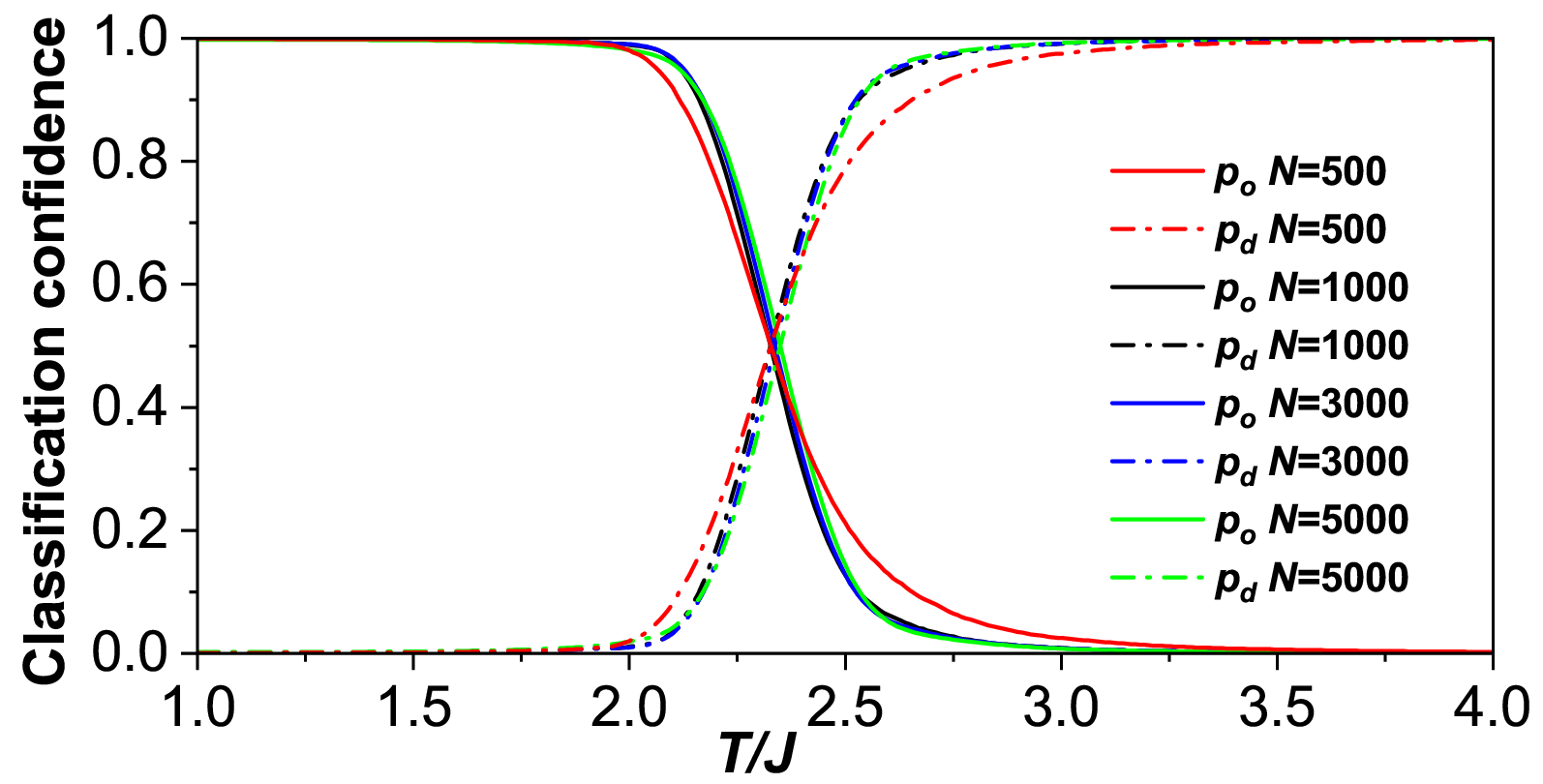}
\begin{tabular}{|c|c|c|c|c|}
\hline
$N$ & $r_c/R$                  & $N_3$ ($w_3$) & $N_4$ ($w_4$) & $N_5$ ($w_5$) \\ \hline
500                              & 0.1813                  & 68 (11.6\%)       & 418 (83.6\%)      & 14(2.8\%)  \\ \hline
1000                             & 0.1298                  & 76 (7.6\%)  & 850 (85\%)  & 74 (7.4\%)  \\ \hline
3000                             & 0.0747                  & 186 (6.2\%) & 2740 (91.3\%) & 74 (2.5\%)  \\ \hline
5000                             & 0.0582                  & 112 (2.2\%) & 4740 (94.8\%) & 148 (3\%)   \\ \hline
\end{tabular}
  \caption{(Top panel) Classification confidences, $p_\text{o}$ and $p_\text{d}$, versus temperature for $(N,r_c/R)=(1000,0.1298)$, (3000,0.0747) and (5000,0.0582). (Bottom panel) Corresponding fractions of sites with different numbers of neighbors.}	\label{MTc2}
 \end{figure}

Next, we vary the number of lattice sites and adjust $r_c$ accordingly to ensure the majority of sites have four neighbors. However, due to the absence of translational and rotational symmetry in the spherical Fibonacci lattice, this adjustment does not necessarily guarantee that lattices with different site numbers share similar geometric structures. This contrasts with the 2D square lattice, where structural consistency is maintained. 
In addition to the previously discussed case of $N=1000$, $r_c/R=0.1298$, we also select $N=500$, $r_c/R=0.1813$; $N=3000$, $r_c/R=0.0747$ and $N=5000$, $r_c/R=0.0582$, ensuring that the fractions of sites with four neighbors in the latter three cases are 83.6\%, 91.3\% and 94.8\%, respectively. 
Interestingly, this suggests that the ``irregular area" gradually diminishes as the effective curvature of the spherical lattice decreases. 
The numerical results are visualized in the top panel of Fig.\ref{MTc2}.
Similarly, we determine the phase transition temperatures using two different methods mentioned before. 
First, by identifying the intersection of the classification confidences, we obtain 
$T_c/J=2.33$, $2.33$, $2.34$ and $2.35$ for $N=500$, $1000$, $3000$ and $5000$, respectively. Second, using linear fitting near $T_c$ (see the bottom panel of Fig.\ref{Tc1}), the corresponding phase transition temperatures are giving
$T_c/J=2.331\pm 0.0742$, $2.331\pm 0.0900$, $2.348\pm 0.0615$ and $2.358 \pm 0.0496$ for the corresponding system sizes. Compared with the results for planar square lattice with a similar number of sites and periodic boundary conditions, as the fraction of sites with four neighbors increases, the $T_c$ of the spherical Ising model gradually approaches that of the planar square lattice.
All these findings suggest that even for a simple system like the Ising model, the spherical Fibonacci lattice exhibits properties distinct from those of planar lattices, similar to the observations in Ref.\cite{SIsing23}.

 \section{Antiferromagnetic Ising model on a spherical Fibonacci lattice}\label{IV}

\begin{figure}[ht]
\centering
  \includegraphics[width=1.108in]{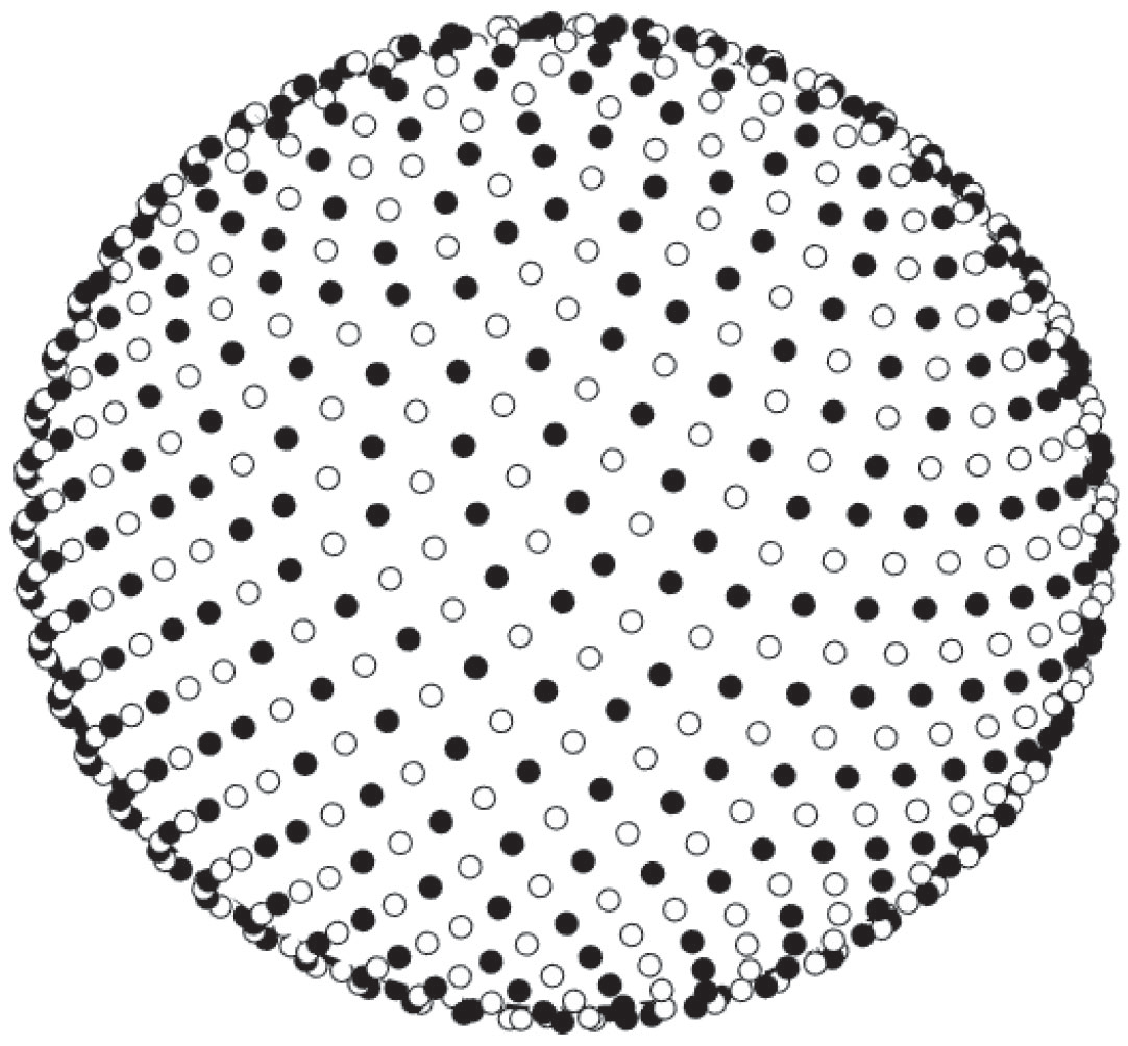}
  \includegraphics[width=1.108in]{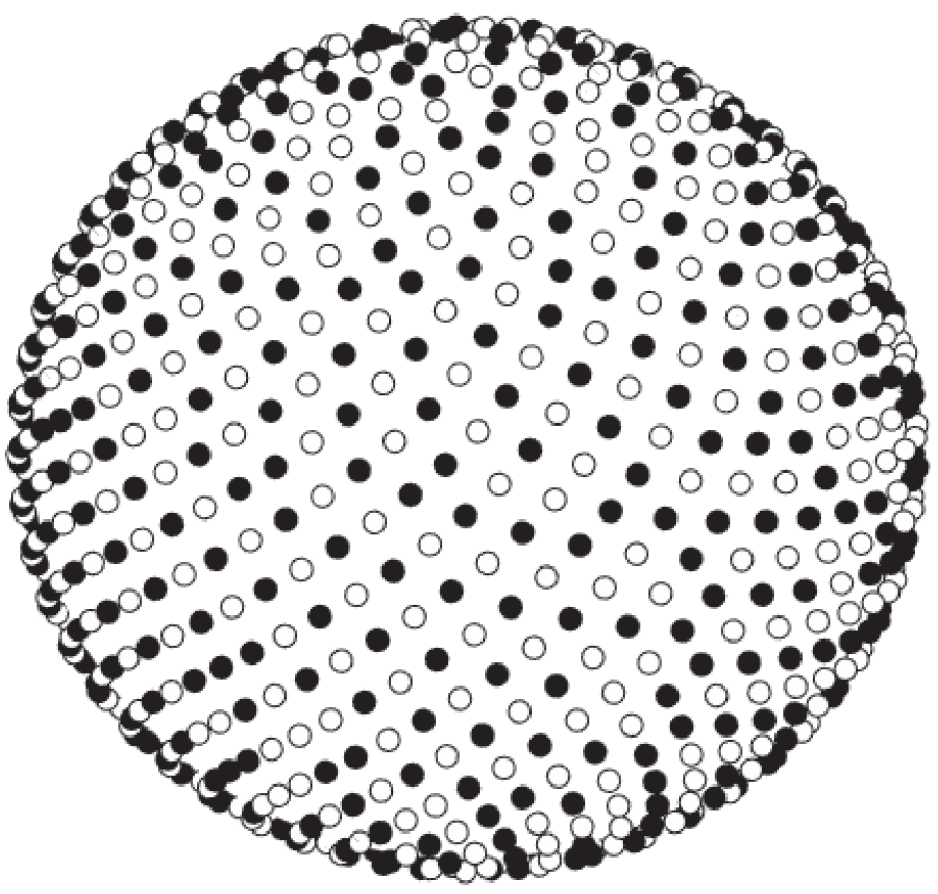}
  \includegraphics[width=1.108in]{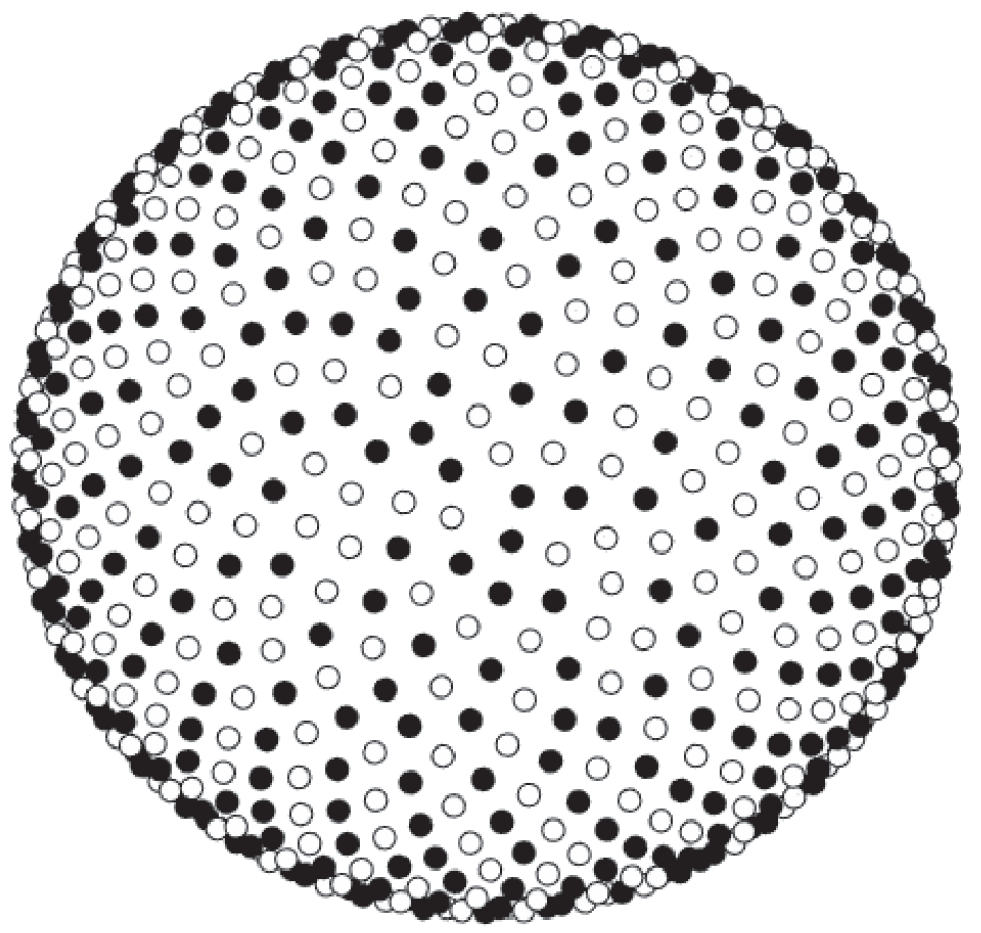}\\
  \includegraphics[width=1.28in]{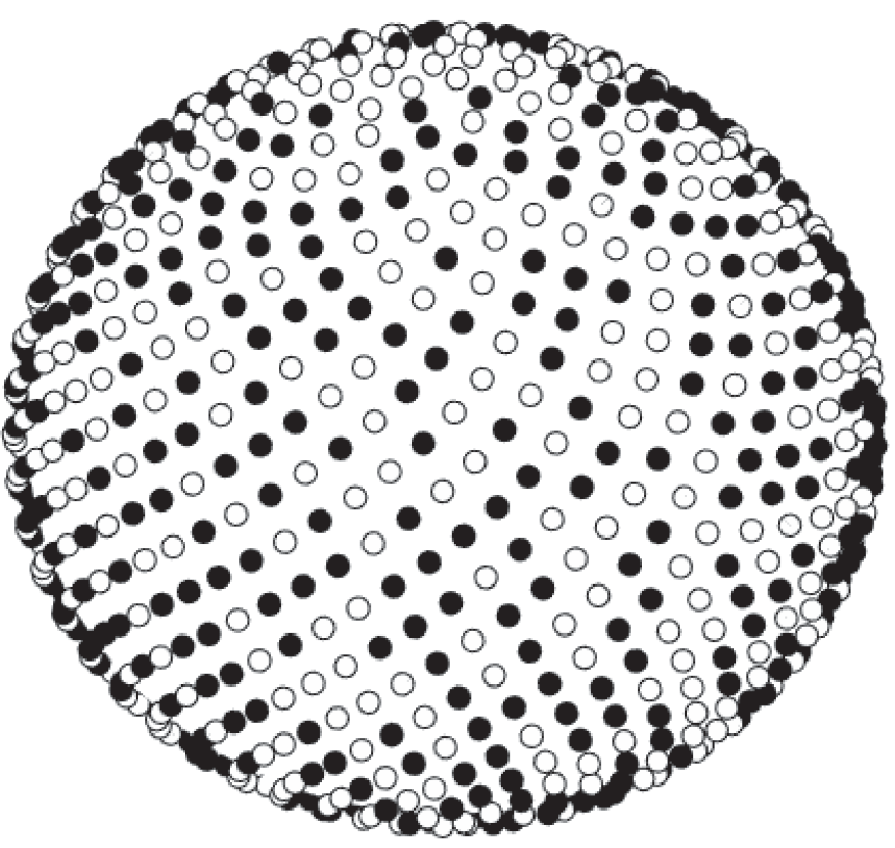}
  \hspace{0.3in}
  \includegraphics[width=1.24in]{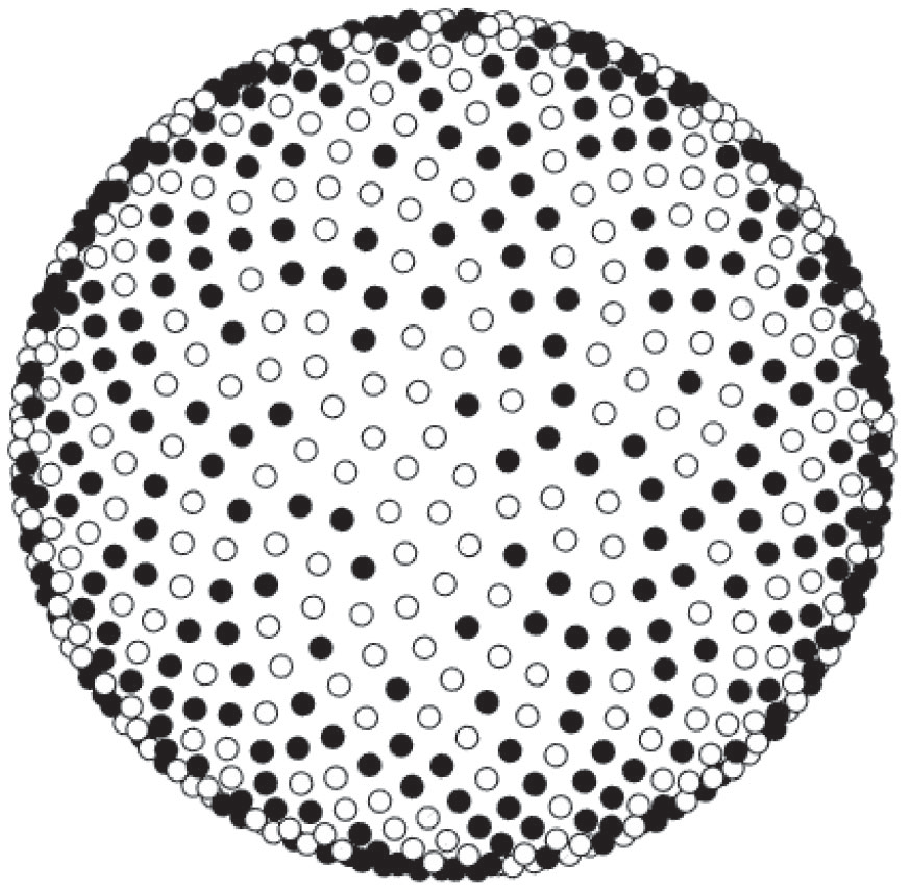}
  \caption{Top row: Spin configurations of the antiferromagnetic Ising model on a spherical Fibonacci lattice at $T/|J|=0.0$ (left) and at $T/|J|=2.0$ from two different directions (middle and right).
  Bottom row: Two perspectives of a stable spin configuration of the same model at $T/|J|=8.0$.
  }	\label{antizerotem}
 \end{figure}

We now consider the case of negative spin coupling, $J<0$. In this scenario, the nearest spins tend to anti-align, leading to the formation of an ordered antiferromagnetic phase. Consequently, the connectivity features of the lattice will influence the ground-state structures. We set $J=-1$ and focus on the spherical Fibonacci lattice with $N=1000$ and $r_c/R=0.1298$. 

In contrast to the ferromagnetic case on the spherical Fibonacci lattice, Monte Carlo simulations at zero temperature reveal that while most neighboring spins align antiparallel, the nonuniform geometry of the spherical lattice, where certain sites have 3 or 5 neighbors, prevents all local interactions from being simultaneously satisfied. As a result, some neighboring spins remain aligned in the same direction. One of the ground-state configuration is shown in the upper-left panel of Fig.~\ref{antizerotem}. This behavior exemplifies the well-known phenomenon of geometric frustration. Geometric frustration refers to situations where, due to geometric constraints of the lattice, local interactions cannot all be simultaneously optimized, preventing the system from establishing a perfectly ordered state \cite{moessner2006geometrical,ronceray2019range}. A classic example is the antiferromagnetic triangular lattice: placing one spin at each vertex of an equilateral triangle with antiferromagnetic interactions, two spins can align antiparallel to satisfy their interaction, but the third spin cannot simultaneously satisfy antiferromagnetic interactions with both neighbors. As a consequence, geometric frustration leads to the existence of a large number of degenerate ground states at zero temperature. In other words, even when the system reaches its lowest energy, it cannot establish a unique globally ordered configuration.
As the temperature increases, while remaining low enough, the spin distribution stays in an ordered state, though it becomes irregular at certain sites. We present the corresponding spin configuration at $T/|J|=2.0$ as viewed from two different directions in the middle and right panels of the top row of Fig.\ref{antizerotem}. As the temperature increases sufficiently, the ordered structure is gradually disrupted, resulting in a disordered state similar to that observed in the ferromagnetic Ising model (shown in the bottom row of Fig.\ref{lowtem}).

For 2D square lattice, the phase transition temperature of the antiferromagnetic Ising model can also be theoretically predicted, $T^\Box_c/J\simeq 2.269$\cite{10.1007/978-3-642-73193-8_19}, which is the same as that of the ferromagnetic Ising model.
To determine $T_c$ for the spherical Fibonacci lattice, we will continue applying the previous methods. 
However, due to the less regular spin patterns and the presence of irregular neighbors on the sphere, the antiferromagnetic model exhibits exotic boundaries even at zero temperature. Therefore, the previous methods for estimating magnetic susceptibility are not applicable in this case. We can only calculate the specific heat by using $C_V=\left(\frac{\partial E}{\partial T}\right)_V$ and Eq.(\ref{Cveqn}).

\begin{figure}[ht]
\centering
  \includegraphics[width=3.1in]{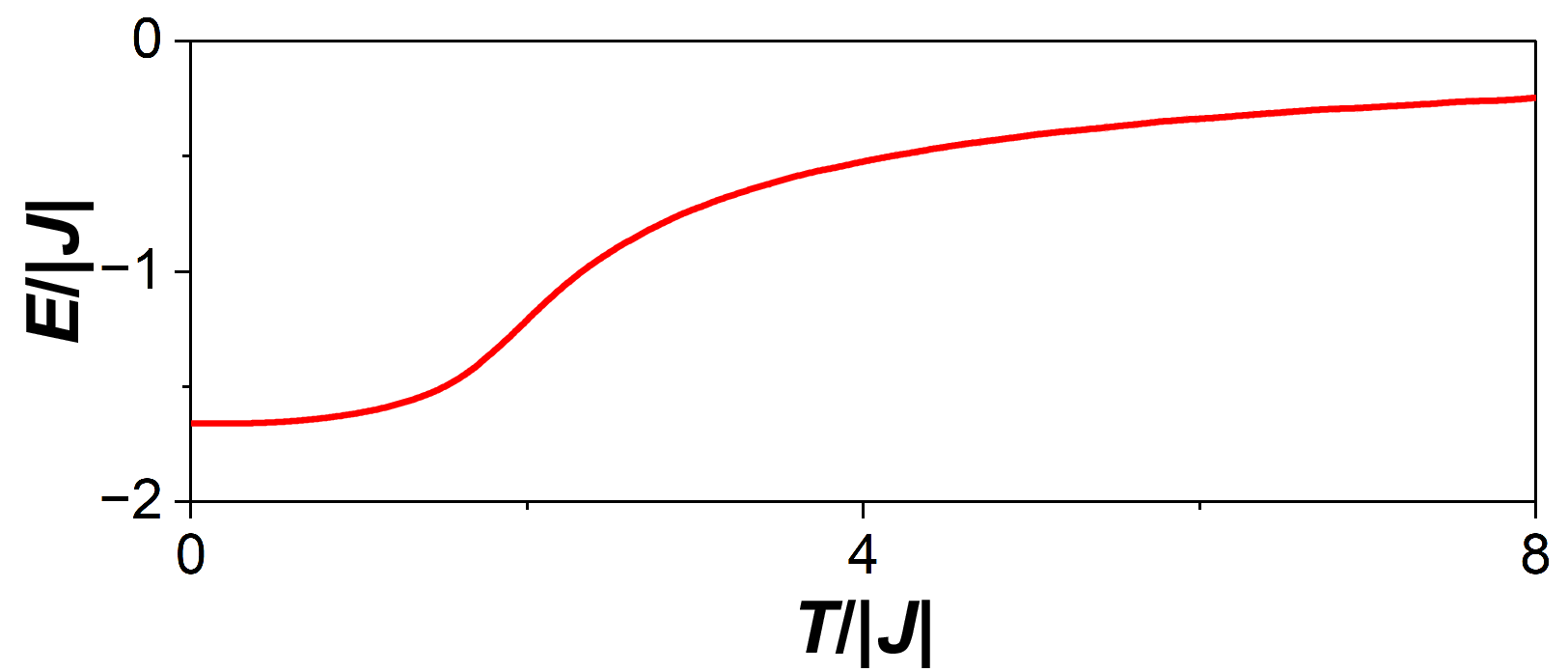} \\
  \includegraphics[width=3.1in]{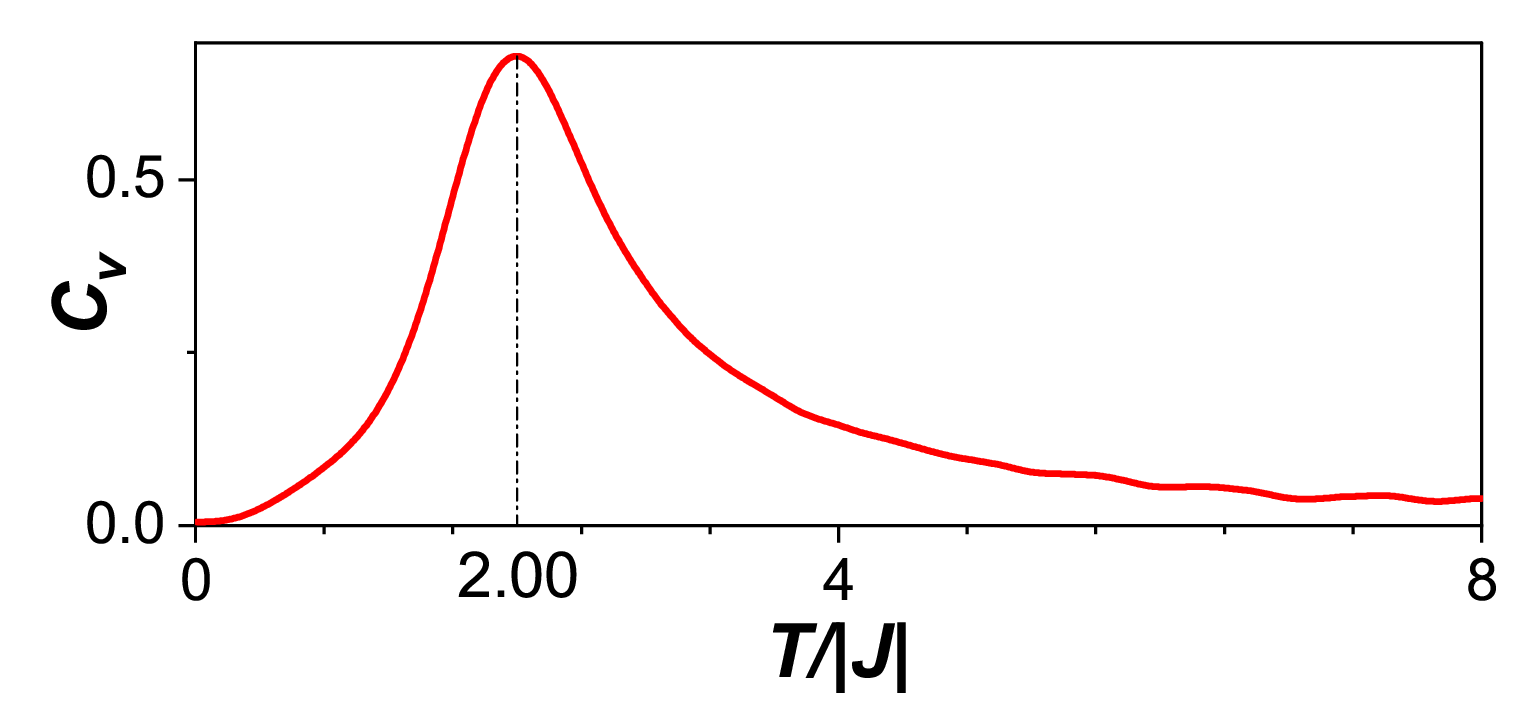}\\
    \includegraphics[width=3.1in]{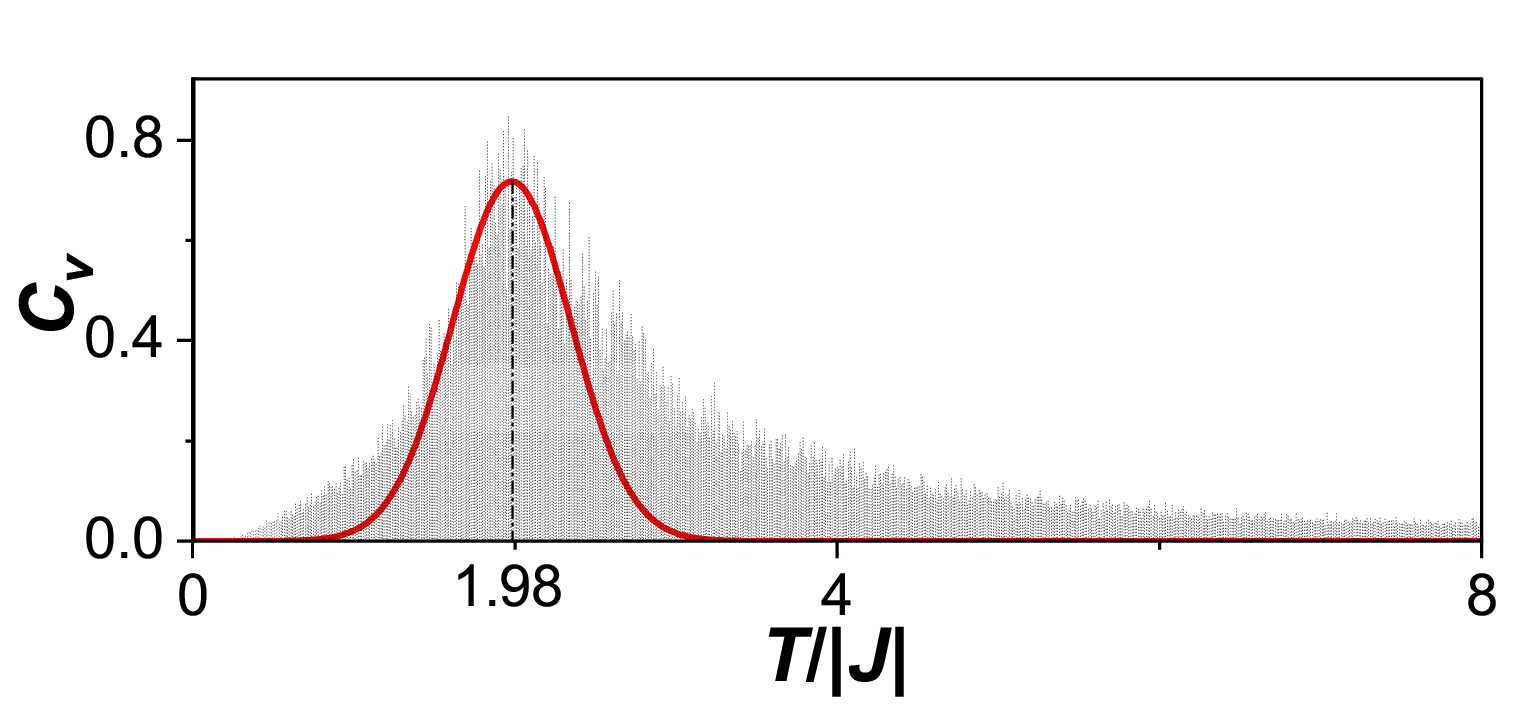}
  \caption{(Top panel) Total energy as a function of temperature for the spherical antiferromagnetic
Ising model with $N=1000$. (Middle panel) Specific heat as a function of temperature for the same model.
(Bottom panel) Specific heat as a function of temperature, calculated using Eq.(\ref{Cveqn}). The black dashed lines represent the raw Monte Carlo data points, and the red solid line shows the Gaussian fit around the peak.}	\label{ACv1}
 \end{figure}

Following the same procedure, we compute the total energy for stable spin configurations at each temperature and plot $E$ vs $T$ in Fig.\ref{ACv1} (top panel). By fitting the $E-T$ curve with spline functions and taking derivatives with respect to $T$, we obtain the specific heat $C_V$, which is also presented in Fig.\ref{ACv1} (middle panel). 

Based on the behavior of $C_V$, we obtain an estimation of the critical temperature: $T_c/|J|=2.00$, which is relatively lower than that of the antiferromagnetic square lattice. This is understandable. As previously mentioned, due to the nonuniform distribution of sites on the spherical lattice, some lattice points have an odd number of neighbors (3 or 5), leading to the phenomenon of geometric frustration. The geometric frustration reduces the system’s tendency to establish global order, thereby lowering the phase transition temperature. After performing the Bootstrap error analysis, the result is refined to $T_c/|J|=1.999\pm 0.006$. To cross check our results, we also employ Eq.(\ref{Cveqn}) to analyze the behavior of $C_V$ and present the numerical results in the bottom panel of Fig.\ref{ACv1}. By fitting the data with a Gaussian function, we estimate the critical temperature to be $T_c/|J|=1.98$. Furthermore, Bootstrap error analysis refines this estimate to $T_c/|J|=1.982\pm 0.012 $.

\begin{figure}[ht]
\centering
  \includegraphics[width=3.1in]{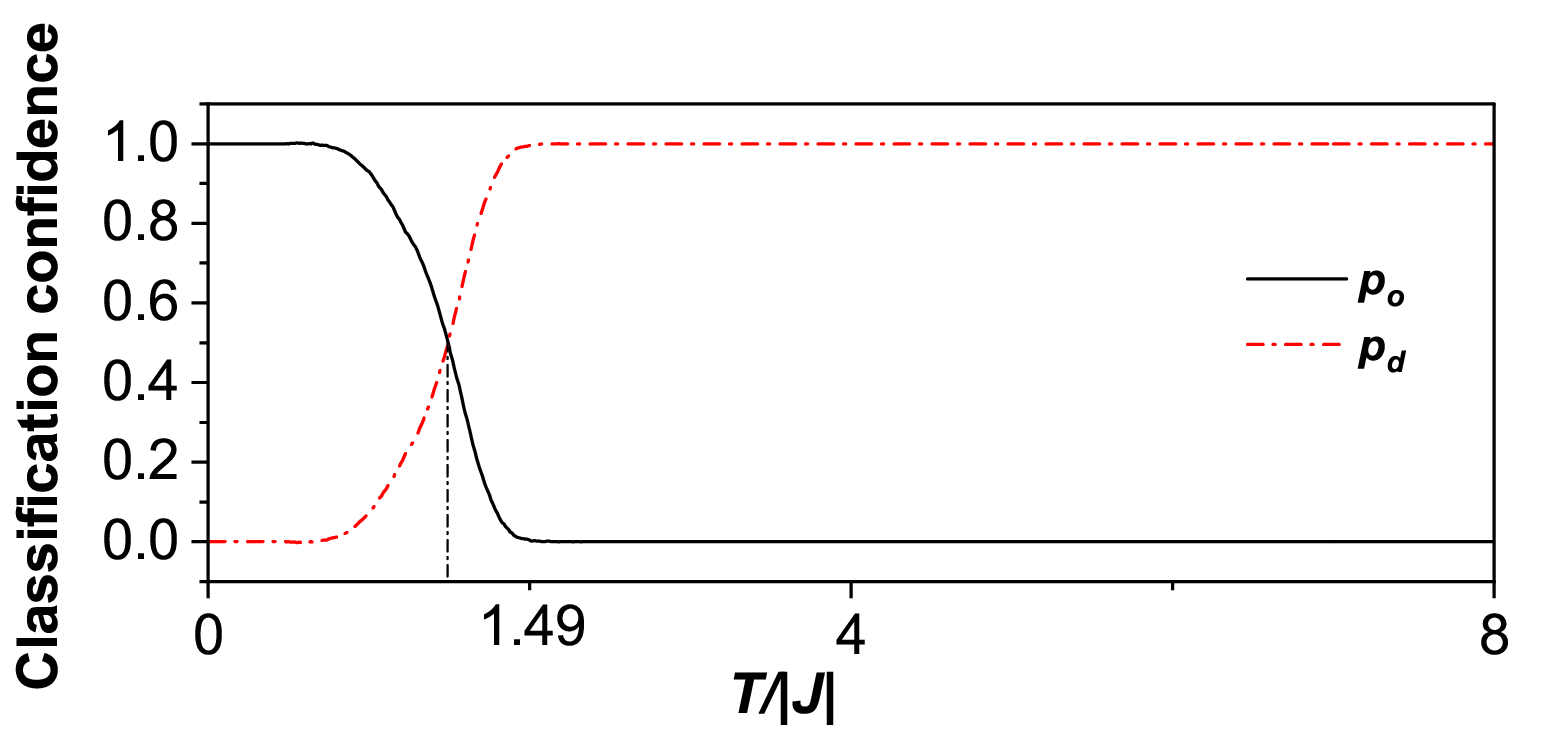} \\
    \includegraphics[width=3.1in]{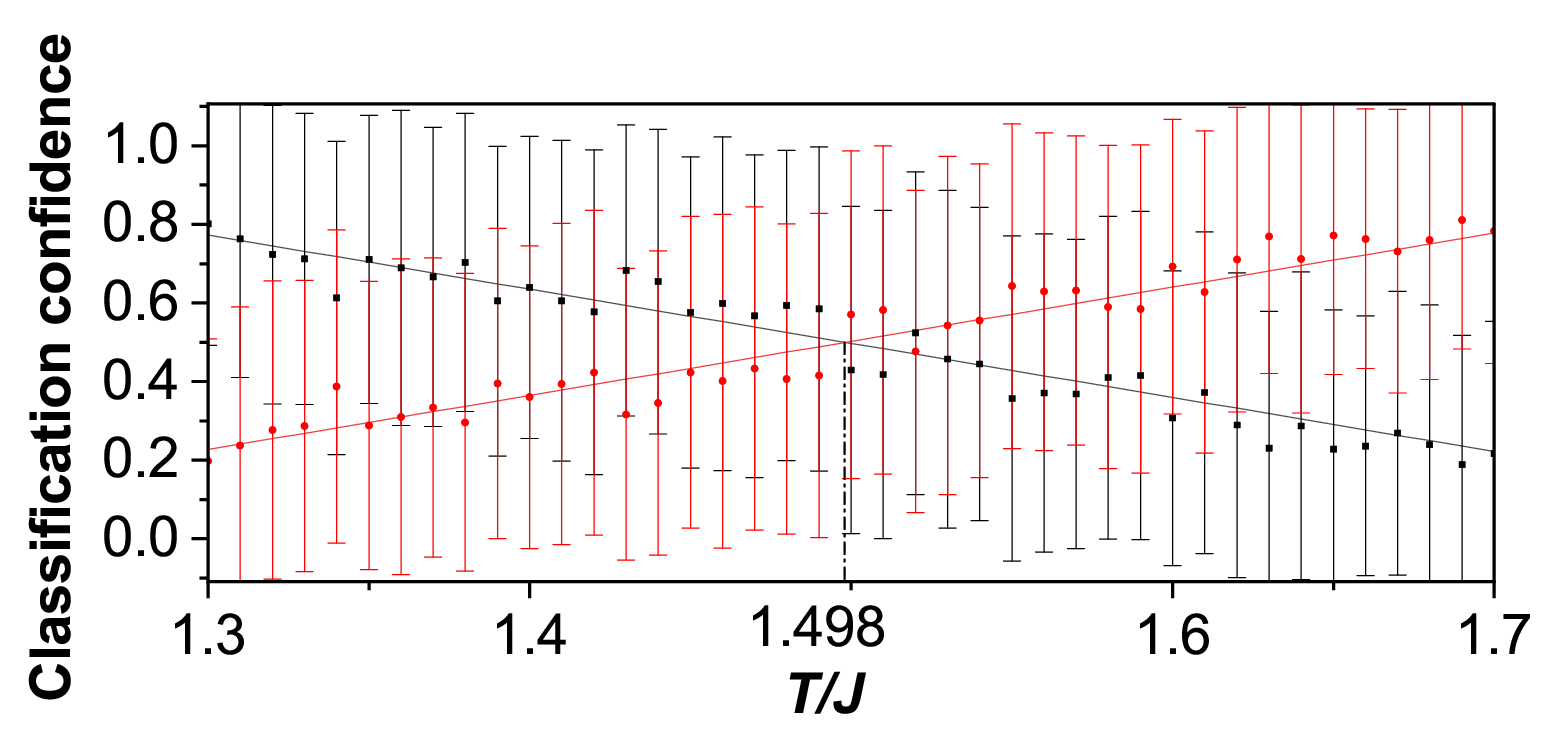}
    \caption{(Top panel) Critical temperature of the antiferromagnetic Ising model on a spherical Fibonacci lattice.
    (Bottom panel) The corresponding error analysis of $T_c$ via linear fitting near the point $p_\text{o}=p_\text{d}.$}\label{antiphase}
 \end{figure}
\begin{figure}[ht]
\centering
  \includegraphics[width=1.5in]{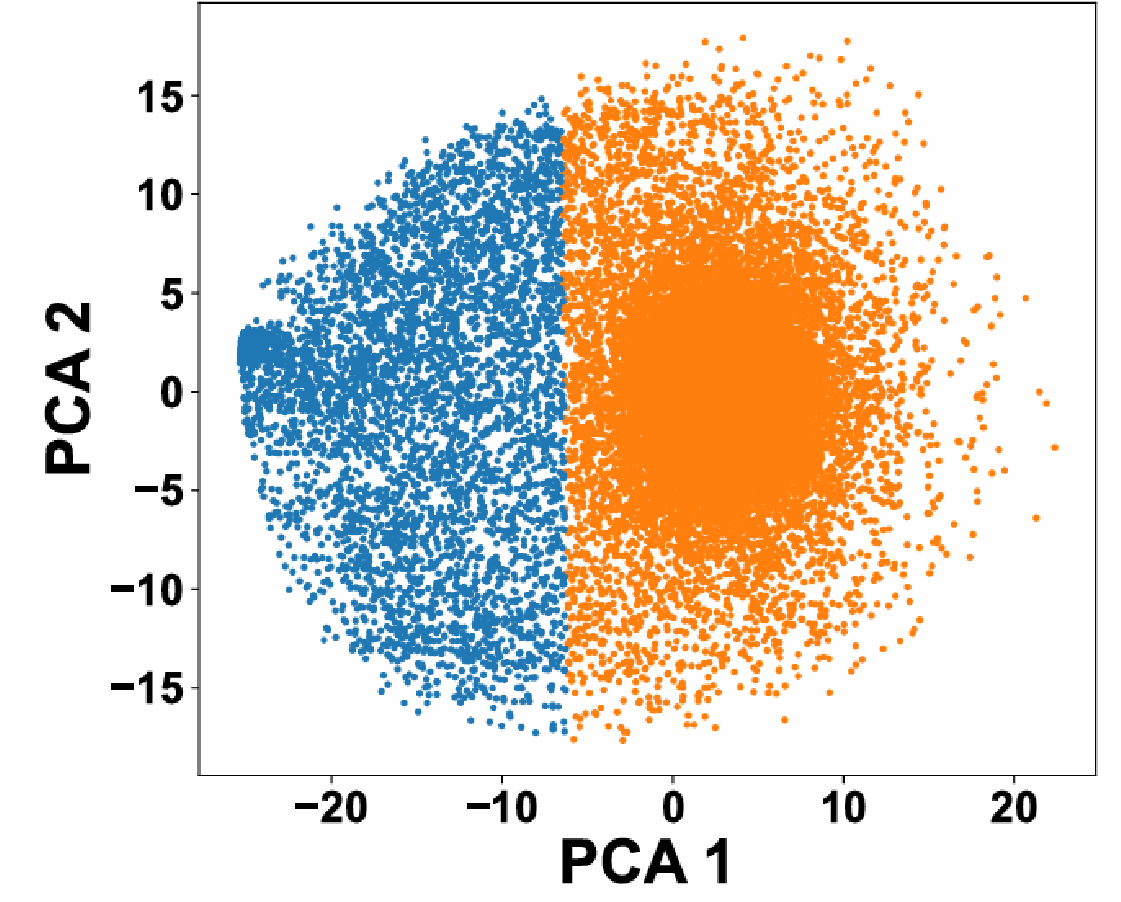}
  \includegraphics[width=1.5in]{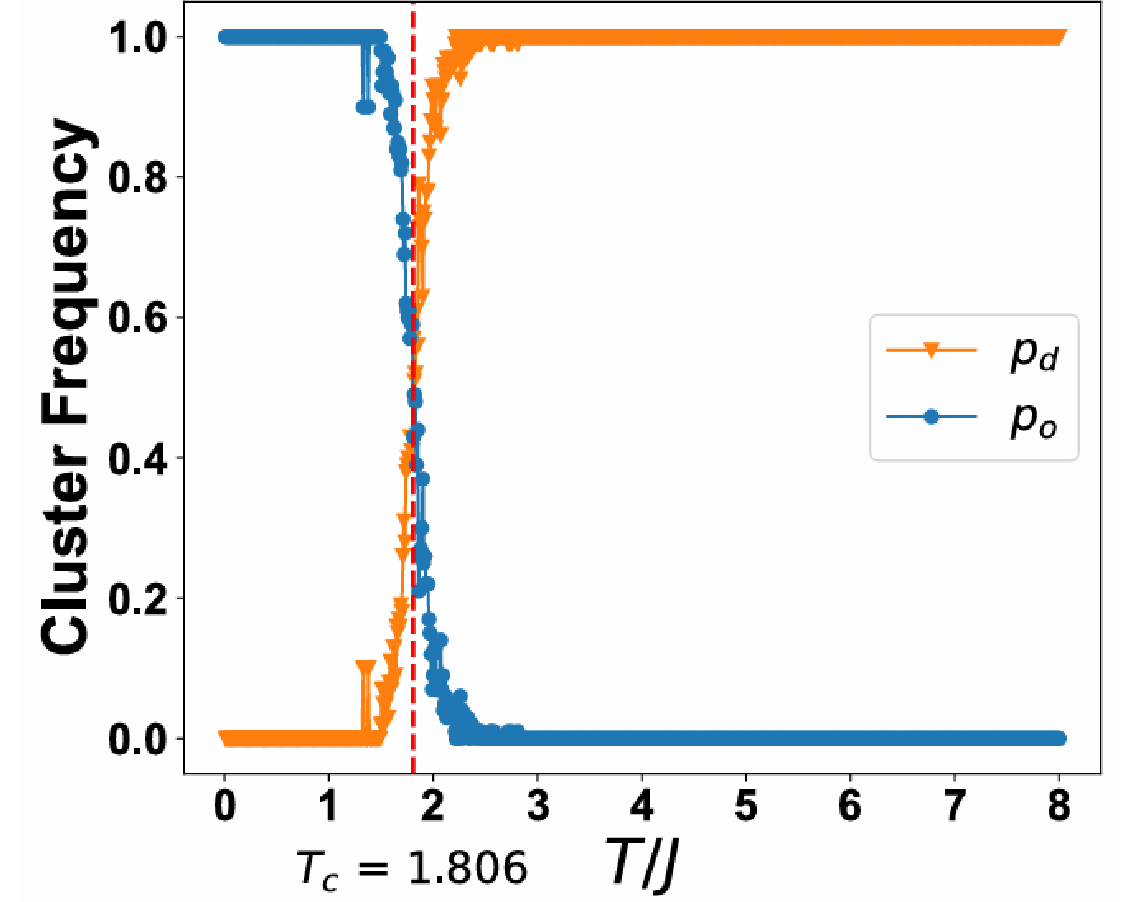} 
  \caption{(Left panel) The similarity of different spin configurations with respect to the two principal components. (Right panel)The cluster fraction, $p_\text{o}$ and $p_\text{d}$, versus temperature of the antiferromagnetic Ising model on a spherical Fibonacci lattice for $N=1000$ .
  }	\label{PCA2}
 \end{figure}

Similar to the spherical ferromagnetic model, we estimated the phase transition temperature of the spherical antiferromagnetic Ising model using both GCNs and unsupervised learning approaches. The numerical results from the GCNs are shown in the top panel of Fig.\ref{antiphase}, yielding a critical temperature of $T_c/|J|=1.49$. In the bottom panel, we present the linear fittings of $p_\text{o}(T)$ and $p_\text{d}(T)$ near $T_c$, with the estimated uncertainty calculated via Eq.(\ref{kerror}) as $T_c/|J|=1.498 \pm 0.210$. The results from unsupervised learning are shown in Fig.\ref{PCA2}: the left panel depicts the clustering outcome, which is unlike the spherical ferromagnetic model and does not show clearly separated clusters; the right panel shows the cluster fractions as a function of temperature, from which the phase transition temperature is estimated to be $T_c/|J|=1.806$.

The estimated $T_c$ is significantly lower than the value obtained from the specific heat $C_V$. This discrepancy can be attributed to two main factors. First, in the antiferromagnetic case, the distinction between the ordered and disordered phases near $T_c$ is less obvious than in the ferromagnetic case, which affects the GCNs’ ability to differentiate the two phases, and the unsupervised learning also fails to produce well-separated clusters. Second, supervised learning requires pre-input of the zero-temperature ordered spin configurations; however, the spherical antiferromagnetic Ising model exhibits geometric frustration, leading to a highly degenerate ground state. As a result, the initial training samples do not necessarily cover all possible degenerate ground states, which reduce the sensitivity of the machine learning approach. 
This discrepancy also reflects a limitation of the machine learning approach: while it performs well for systems with clear phase boundaries (such as the ferromagnetic case), its estimates of $T_c$ in frustrated antiferromagnetic systems are less reliable. Even unsupervised learning struggles to generate well-separated clusters, which further indicates that geometric frustration reduces the ability of machine learning to capture the underlying phase transition.
 Therefore, we consider the $T_c$ obtained from the specific heat method to be more reliable. 

\section{Conclusion}\label{V}
In this paper, we studied the Ising model on a spherical Fibonacci lattice, a structure that balances relative uniformity with irregular sites that significantly influence spin behavior. Employing Monte Carlo simulations, principal component analysis (PCA) and graph convolutional networks (GCNs), we analyzed spin configurations and identified phase transition temperatures for both ferromagnetic and antiferromagnetic cases. 

In the ferromagnetic scenario, sites with fewer neighbors were more prone to spin flips at low temperatures, an effect that weakened as temperature rose. This behavior led to a phase transition temperature higher than that of the planar Ising model, highlighting how the lattice's curvature and connectivity bolster ferromagnetic order. Conversely, in the antiferromagnetic case, lattice irregularities induces geometric frustration, which reveals the complex interplay between geometry and spin interactions in non-planar systems. These findings underscore the pivotal role of geometric features in shaping spin dynamics and phase transitions, distinguishing this system from traditional planar models.

This work enhances our understanding of the Ising model in non-planar geometries and holds particular relevance for spin systems in unique settings, such as microgravity environments, where planar assumptions may not apply. By elucidating the effects of curvature and connectivity on spin interactions, our study paves the way for future theoretical and experimental investigations into the behavior of spin systems on complex lattice structures.

While our study focuses on the Ising and XY models on spherical Fibonacci lattices, the GCNs is broadly applicable to other statistical models and non-Euclidean geometries. For instance, multi-state systems such as the $q$-state Potts model can be naturally incorporated by extending the binary classification task (ordered vs. disordered) to a multi-class task (distinguishing $q$ states), allowing GCNs to effectively identify complex phase diagrams across multiple competing states. Prior to performing the multi-class task, unsupervised learning methods can be used to extractfeatures of spin configurations to better distinguish between different phases.

Beyond spherical lattices, GCNs are well suited for systems defined on other curved geometries. For example,geodesic domes whose triangular mesh structure locally resembles that of Fibonacci lattices, while toroidal embeddings introduce periodic boundary conditions that can be encoded in the adjacency matrix. These features enable the GCNs to capture long-range correlations and topological constraints intrinsic to the geometry.

Finally, combining GCNs with unsupervised learning techniques such as PCA provides a promising approach to uncover intrinsic phase structures in complex systems, helping to reveal subtle correlations that may be difficult to detect with purely supervised learning. Overall, these extensions demonstrate the versatility of the GCNs framework in studying a wide range of statistical systems on non-Euclidean geometries.

\section{Acknowledgments}
H.G. was supported by the Innovation Program for Quantum Science and Technology-National Science and Technology Major Project (Grant No. 2021ZD0301904) and the National Natural Science Foundation of China (Grant No. 12074064). X. Y. H. was supported by the the National Natural Science
Foundation of China (Grant No. 12405008) and the Jiangsu Funding Program for Excellent Postdoctoral Talent (Grant No. 2023ZB611).

\appendix

\section{Bootstrap Error Analysis Method}\label{app1}
To assess the statistical uncertainty of the estimation of $T_c$ based on the behavior of $C_V$, we employ the nonparametric Bootstrap resampling method. This technique simulates statistical fluctuations in the sample distribution without assuming that the data follows a specific distribution, making it well-suited for error estimation in finite-sized systems. Using Eq.(\ref{Cveqn}) as an example, the detailed steps are as follows:
\begin{enumerate}
    \item \textbf{Data Preparation} \\
       Suppose the system undergoes Monte Carlo simulations at temperature points $\{T_i\}_{i=1}^N$. At each temperature point, $M$ independent energy observations are collected, forming the dataset $\mathcal{D} = \{E_{i1}, E_{i2}, \dots, E_{iM}\}_{i=1}^N$.
    \[
    \mathcal{D} = \{E_{i1}, E_{i2}, \dots, E_{iM}\}_{i=1}^N.
    \]
 \item \textbf{Resampling Generation} \\
    For each temperature point $T_i$, perform random sampling with replacement:
    \[
    \mathcal{D}^{(b)}_i = \left\{E_{ik}^{(b)} \mid k \in \{1,2,\dots,M\}, \quad E_{ik}^{(b)} \sim \mathcal{D}_i \right\},
    \]
    where the superscript $(b)$ denotes the $b$-th Bootstrap iteration (with a total of $B=1000$ iterations).
\item \textbf{Statistic Calculation} \\
    For each Bootstrap sample $\mathcal{D}^{(b)}$, compute the target statistics:
    \begin{itemize}
        \item \textbf{Specific Heat Capacity}:
        \[
        C^{(b)}(T_i) = \frac{\langle E^2 \rangle^{(b)}_i - \langle E \rangle^{(b)2}_i}{ T_i^2}.
        \]
        \item \textbf{Critical Temperature $T_c^{(b)}$}: Fit the $C^{(b)}(T)$ curve with a Gaussian function and locate the peak temperature.
    \end{itemize}
\item \textbf{Error Estimation} \\
    \begin{itemize}
        \item \textbf{Standard Deviation}:
        \[
        \sigma_{T_c} = \sqrt{\frac{1}{B-1} \sum_{b=1}^B \left(T_c^{(b)} - \bar{T}_c\right)^2}, \quad \bar{T}_c = \frac{1}{B}\sum_{b=1}^B T_c^{(b)}.
        \]
        \item \textbf{Confidence Interval}: Take the 2.5\% and 97.5\% percentiles of $\{T_c^{(b)}\}$ to define the 95\% confidence interval $[T_c^{\text{low}}, T_c^{\text{high}}]$.
    \end{itemize}
\end{enumerate}

\section{Methodology of phase classification: Details of GCNs}\label{app3}

In this paper, we use the spherical Fibonacci lattice as a sample, where the lattice information is represented as a graph $\mathcal{G}$. All relevant data is stored in the degree matrix $\boldsymbol{D}$ and the adjacency matrix $\mathcal{A}$. The convolution operation is performed using the Laplacian matrix
$\boldsymbol{L}=\boldsymbol{D}-\mathcal{A}$.

The features of the lattice points are organized into the feature matrix
$\boldsymbol{X}=\left(\boldsymbol{s}_{1}, \cdots, \boldsymbol{s}_{N}\right)^{T} \in \mathbb{R}^{N}$, where $N$ is the number of sites and $\boldsymbol{s}_{i}$ denotes the spin at lattice site $i$. We then apply the Random Walk normalized Laplacian
$\boldsymbol{L}^{\mathrm{rm}}=\boldsymbol{D}^{-1} \boldsymbol{L}$ for feature extraction, combining it with the feature matrix to obtain:
\begin{equation}\label{e6}
\boldsymbol{H}=\operatorname{ReLu}\left(\boldsymbol{L}^{r m} \boldsymbol{X} \boldsymbol{W}_{h}+\boldsymbol{b}_{h}\right)
\end{equation}
Here, $\operatorname{ReLu}$ serves as the activation function,
$\boldsymbol{W}_{h}\in \mathbb{R}^{1\times1}$
  is the weight, and
$\boldsymbol{b}_{h}\in \mathbb{R}^{N\times1}$
  is the bias. Finally, we employ a fully connected layer along with the $\emph{softmax}$ function to aggregate the hidden layer, resulting in an output
$\boldsymbol{H} \in \mathbb{R}^{N \times 1}$
  that generates classification confidence for the ordered and disordered phases, denoted as
$p_\text{o}$  and $p_\text{d}$. The temperature at which
$p_\text{o}=p_\text{d}$ defines the phase transition temperature $T_c$.

\end{document}